\documentclass[preprint,preprintnumbers,nofootinbib,floatfix]{revtex4}
\usepackage{times,bm,bbm,bbold,amssymb,amsbsy,amsthm,amsmath,amsfonts,mathrsfs,graphics,graphicx,color,xcolor,multirow}

\definecolor{myurlcolor}{rgb}{0,0,0.7}
\definecolor{myrefcolor}{rgb}{0.8,0,0}
\usepackage{hyperref}
\hypersetup{colorlinks,linkcolor=myrefcolor,citecolor=myurlcolor,urlcolor=myurlcolor}

\begin{document}

\begin{titlepage}
\vspace*{10mm}
\begin{center}
{\Large \bf Almost-quantum correlations violate the isotropy and homogeneity principles in flat space}

\vspace*{10mm}

\large{Akbar Fahmi} \footnote{fahmi@theory.ipm.ac.ir}

\vspace*{.1cm}

\small{(\today)}

\vspace*{.5cm}





{\bf Abstract}
\end{center}
One of fascinating phenomena of nature is quantum nonlocality, which is observed upon measurements on spacelike entangled systems.
However, there are sets of post-quantum models which have stronger correlations than quantum mechanics, wherein instantaneous communication
remains impossible. The set of ``almost quantum correlations" is one of post-quantum models which satisfies all kinematic axioms of
standard quantum correlations except one, meanwhile they contain correlations slightly stronger than quantum correlations.
There arises the natural question whether there is some fundamental principle of nature which can genuinely characterizes quantum correlations.
Here, we provide an answer and close this gap by invoking the ``isotropy and homogeneity" principles of the flat space as a conclusive and distinguishing criterion to rule out the almost-quantum correlations model. In particular, to characterize quantum correlations we impose the isotropy and homogeneity symmetry group structure on the almost quantum correlations model and request that the joint probability distributions corresponding to the Born rule remain invariant. We prove that this condition is sufficient (and necessary) to reduce the almost quantum correlations model to quantum mechanics in both bipartite and multipartite systems.



\end{titlepage}

\section*{}

\emph{Introduction}.---Nonlocality property of nature is one of fundamental discoveries in recent decades. Nonlocality was first predicted in the remarkable Einstein-Podolsky-Rosen theorem (EPR) \cite{EPR} and later was manifested in the Bell-type inequalities \cite{Bell,Bell1,CHSH}. Quantum mechanics violates these inequalities, which is also strongly supported by a wide range of experimental tests \cite{Han,Sha, Ant,Win,Pan,Sup}. These results indicate strong evidence against local hidden variable theories \cite{Bell1}. It has been known that the maximum \emph{quantum} nonlocality is still weaker than what the no-signaling principle allows \cite{PR1}. Notably, there are abstract and deterministic nonlocal box models \cite{PR1} which allow even stronger than any correlations realizable in quantum mechanics, wherein still instantaneous communication remains impossible and the no-signalling principle is respected. Specifically, nonlocal boxes violate the Clauser-Horne-Shimony-Holt (CHSH) inequality \cite{CHSH} by the algebraic maximum value $4$ \cite{PR1,BCP,JM,BP1,BP2,Mas}, whereas quantum correlations achieve at most $2\sqrt{2}$ (Tsirelson's bound) \cite{Tsi}.

This propounds the question whether there is a fundamental principle which can rule out the post-quantum models conclusively. In this respect, numerous attempts have been down to search for theory-independent physical principles which some how characterize quantum correlations. For example, several information-theoretic principles such as nontrivial communication complexity \cite{Dam,Bra2}, no advantage for nonlocal computation \cite{NC}, information causality \cite{IC} and its generalization \cite{IC2,IC3}, macroscopic locality \cite{ML,ML1} and its refined version many-box locality \cite{MBL}, and local orthogonality \cite{LO} were proposed as revelent principles to provide characteristic bounds on the set of post-quantum correlations. Although these principles are not able to recover \emph{quantum} correlations, their analysis can be helpful in order to lead us to ward identifying the more fundamental and sufficient physical principle for quantum correlations.



In addition to the abstract nonlocal models \cite{PR1,BCP,JM,BP1,BP2,Mas}, there is also another type of post-quantum models which is refereed to the ``almost quantum correlations" (AQCs) model \cite{Alm}. The set of AQCs satisfies all kinematic postulates of standard quantum correlations except one: the postulate of commutation of two measurement operators for two different parties is replaced by the equality of arbitrary permutations of the measurement operators on a specific normalized vector state. Evidently, the set of AQCs is strictly larger than the standard quantum set and violates the bound maximum quantum correlations. In addition, the AQCs set satisfies most of aforementioned information-theoretic principles such as the nontrivial communication complexity \cite{Dam,Bra2}, the no advantage for nonlocal computation \cite{NC}, the macroscopic locality principle \cite{ML,ML1} and the many-box locality \cite{MBL}, and the local orthogonality \cite{LO} principles. furthermore, there is numerical evidence which the AQCs also satisfy the information causality principle \cite{IC}.

Recently, the no-restriction hypothesis \cite{Alm1} was specifically proposed to distinguish the standard quantum correlations from the AQCs. This hypothesis implies that any mathematically well-defined measurement in generalized probabilistic theory should also be physically allowed. Unlike standard quantum mechanics, a generalized probabilistic theory satisfying this hypothesis cannot reproduce the set of AQCs \cite{Alm1}. In a similar fashion, Specker principle \cite{Sp} has been reformulated to apply as a device-independent witness of the AQCs such that ``if in a set of measurements every pair is compatible, then all the measurements are compatible." Specifically, unlike quantum correlations, in any general probabilistic theory the structure of measurements reproducing the set of AQCs model contradicts the Specker principle \cite{Alm3}. Thus, it may seem that the no-restriction hypothesis or the Specker principle can be considered as fundamental principles to single out the set of quantum correlations from among other nonsignaling models.


According to above statements, there are several types of principles \cite{Dam,Bra2,NC,ML,ML1,IC,IC2,IC3,MBL,LO,Alm1,Alm3} which are mainly proposed to rule out \emph{specific} post-quantum models \cite{PR1,BCP,JM,BP1,BP2,Mas,Alm}. These information-theoretic principles are designed to detect bipartite post-quantum models \cite{Dam,Bra2,NC,ML,ML1,IC,IC2,IC3,MBL,LO,Alm1,Alm3}, however, these principles fail to distinguish the multipartite post-quantum models from multipartite quantum correlations \cite{NIP,NIP1}. The AQCs model highlights this problem by proposing tripartite version of AQCs model which meets all bipartite information-theoretic principles \cite{Alm}. Unfortunately, the no-restriction hypothesis and the Specker principle do not offer any hint to single out multipartite quantum systems from multipartite AQCs model. In addition, these principles do not address how the AQCs model reduce to standard quantum mechanics in bipartite systems. And, also some multipartite Bell-types inequalities has been employed to witness beyond-quantum-theory AQCs model beyond quantum theory \cite{MuBell,MuBell1}, a sufficient exact characterization of the set of multipartite quantum correlations has remained inaccessible. Thus, one still may wish to search for a principle which is both applicable to multipartite systems and also can delineate how \emph{all} post-quantum models reduce to standard quantum mechanics.


Here, we propose a solution to the above problems by invoking symmetries of the flat space as the natural and universal principles in order to single out quantum correlations conclusively. The flat space is recognized by the isotropy and homogeneity symmetries, two elementary properties which universally hold for all  physical theories. The flat space is usually considered as fixed the background for every experimental setup, and the isotropy/homogeneity symmetry implied that all orientations/locations of an experiment are physically equivalent \cite{Jam}. Specifically, if all elements of an experiment (systems, sources, channels, or other measurement settings of the lab) are rotated/translated equally by the same amount, such transformations do not have any impact on the value of any observable quantities. These principles are independent from the content physical theories and are expected by all observable and experimental measurements \cite{MM,MM1,MM2,LoM1,LoM3,LoM4,LoM5,LoM6,LoM7,LoM8,LoMM}.

In Bell-like scenarios, the relevant observable quantities are probability distributions (correlation functions) of joint measurements of qudits, which should be invariant under the space isotropy/homogeneity symmetry transformations. This property has been demonstrated by a recent ``electronic analogue" of Michelson-Morley experiment \cite{LoMM}. In this experiment, energy dispersion relation of two entangled electrons, which are bound inside a pair of calcium ions, undergo a Michelson-Morley type scenario \cite{MM} to detect anisotropy of the space. It is notable that the isotropy of the electron's dispersion relation was verified at the level of one part in $10^{18}$, which strongly guarantees the invariance of the probability distributions of the entangled quantum system under the flat-space symmetry transformations.



We anticipate that the isotropy/homogeneity principle can also be pivotal in characterizing quantum correlations in the sense that all no-signaling theories which predict stronger correlations than quantum mechanics violate this principle of the flat space. In this regard and in particular, here we prove that the AQCs \cite{Alm} violate the isotropy/homogeneity symmetry principle. To do so, we impose the standard representations of the isotropy/homogeneity symmetry group transformations on the AQCs model \cite{Alm} and request that the probability distributions (the Born rule) remain invariant. The show that this condition is sufficient to reduce the AQCs model to standard quantum mechanics. Moreover, unlike the information-theoretic principles described earlier \cite{Dam,Bra2,NC,ML,ML1,IC,IC2,IC3,MBL,LO,Alm1,Alm3}, we show that the number of parties, dimension of input spaces and dimension of output spaces do not play any role in our framework. This utility opens up a new approach to distinguishing multipartite \emph{quantum} correlations from multipartite \emph{post-quantum} correlations. Specifically, the isotropy/homogeneity principle rules out the AQCs model for multipartite systems with arbitrary dimension of input/output space or every extension thereof \cite{Alm}. It is expected that the impact of the isotropy/homogeneity principle goes beyond merely ruling out the AQCs model. In fact, it seems to us this principle may even apply to the Boolean version of post-quantum models (introduced in Refs. \cite{PR1,BCP,JM,BP1,BP2,Mas}) in the abstract nonlocal box framework, and it also may determine the maximum degree of nonlocality in such models \cite{NLB}.

\begin{figure}
\centering
\includegraphics[scale=.7]{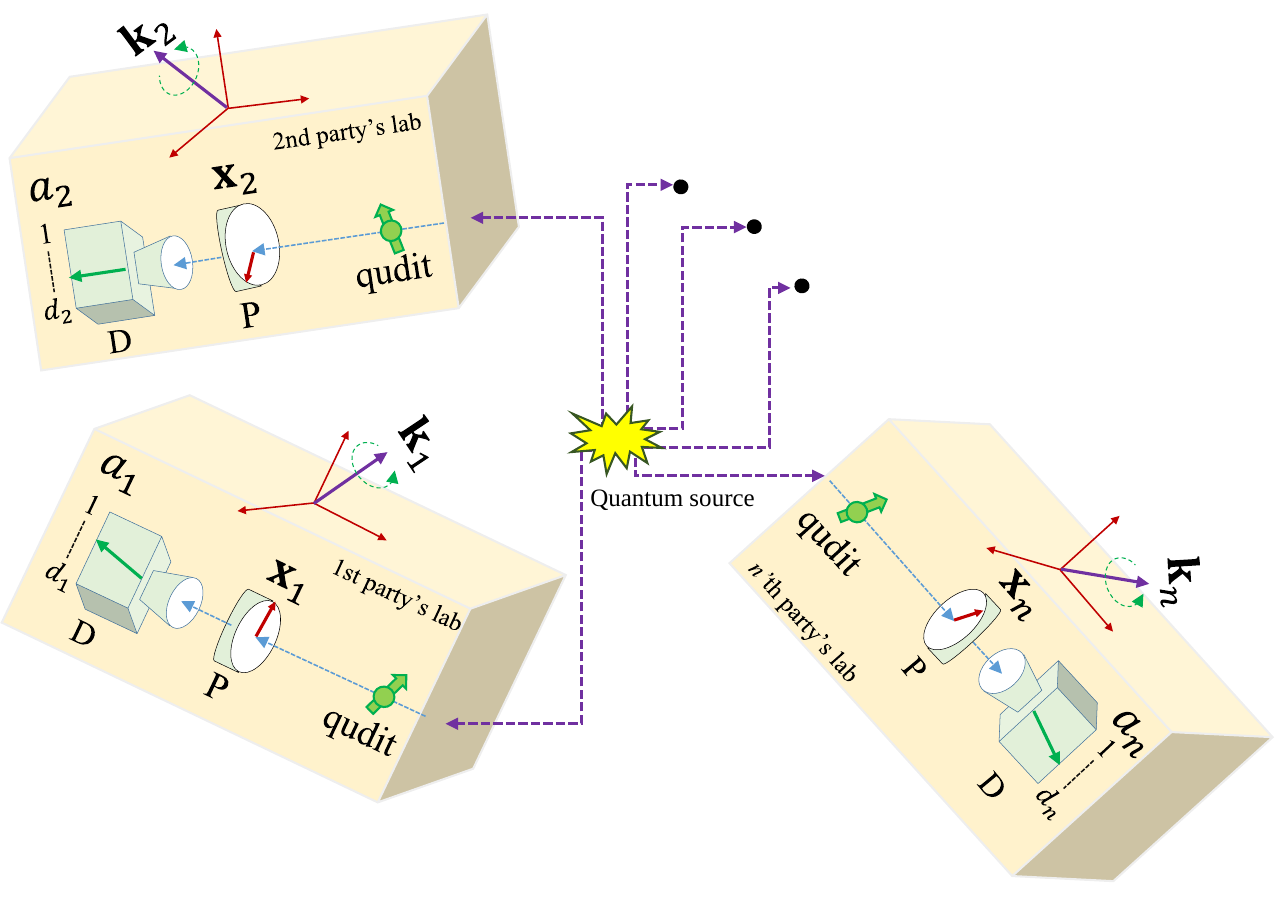}
\caption{Schematic presentation of the local rotation/translation symmetry transformations in the Bell-type scenario for a set of $n$-partite spacelike systems.
After sharing qudits from a quantum source among $n$ parties, the $i$th party ($i=1,\cdots,n$) chooses one of the inputs $\mathbf{x}_{i}\in\left\{1,\cdots,m_{i}\right\}$ and obtains one of the outcomes $a_{i}\in\left\{1,\cdots,d_{i}\right\}$ by performing a measurement on the received qudit to his lab. The isotropy/homogeneity of the space are manifested by rotation/translation of all elements of each experimental setup for each party, including his qudit, a set up other apparatuses followed by a measurement device (P) on the direction $\mathbf{x}_{i}$, the detector (D) which register the result of measurement, and the whole lab by the same amount. We represent the new measurement direction of the $i$th lab with $\mathbf{x}'_{i}$, which is given by an arbitrary rotation or an arbitrary translation about an arbitrary direction
$\mathbf{k}_{i}$ by the value $\theta_{i}$ ($i=1,\cdots,n$). Within this context isotropy/homogeneity symmetry implies that
the joint probability distributions $P(a_{1},a_{2},\cdots,a_{n}|\mathbf{x}_{1},\mathbf{x}_{2},\cdots,\mathbf{x}_{n})=
P(a_{1},a_{2},\cdots,a_{n}|\mathbf{x}'_{1},\mathbf{x}'_{2},\cdots,\mathbf{x}'_{n})$ and any subset of the marginal probability distributions $\sum_{a_{i},\cdots,a_{j}} P(a_{1},a_{2},\cdots,a_{n}|\mathbf{x}_{1},\mathbf{x}_{2},\cdots,\mathbf{x}_{n})=
\sum_{a_{i},\cdots,a_{j}} P(a_{1},a_{2},\cdots,a_{n}|\mathbf{x}'_{1},\mathbf{x}'_{2},\cdots,\mathbf{x}'_{n})$ remain invariant, where the summation has been taken on every arbitrary subset of measurement outputs.
}\label{fig}
\end{figure}

In multipartite systems, such as Bell-like experiments, the \emph{local} isotropy/homogeneity symmetry transformations are defined by arbitrarily rotating/translating every one of subsystems and its measurement apparatuses by the same values such that all joint probability distributions and any subset of marginal probability distributions remain invariant---see Fig. \ref{fig}.

Now we first briefly review kinematic axioms of standard quantum mechanics and the AQCs model and next, we introduce the isotropy/homogeneity symmetry transformations in them. We show that the isotropy/homogenuty symmetry transformations applied on the AQCs model indeed generate infinite classes of other AQCs models whose compatibility with each other will reduce the AQCs models to standard quantum mechanics.

$\vspace{.1cm}$

\emph{Standard quantum mechanics}.---We now review in a formal manner the Bell-type scenario for a set of $n$ spacelike parties. Each party has access to a subsystem and chooses one of $m_{i}$ different inputs to perform measurement on his subsystem and obtains one of the $d_{i}$ possible outputs, as explained in Fig. \ref{fig}. In this scenario, the inputs and the measurement outputs are denoted by $\mathbf{x}_{i}\in\left\{1,\cdots,m_{i}\right\}$ and $a_{i}\in\left\{1,\cdots,d_{i}\right\}$ for each party $i\in\left\{1,\cdots,n\right\}$, respectively. The parties perform measurements on many independent copies of the shared system and together estimate the conditional probability distribution $P(a_{1},a_{2},\cdots,a_{n}|\mathbf{x}_{1},\mathbf{x}_{2},\cdots,\mathbf{x}_{n};|\psi\rangle)$, which is given by the Born rule as
\begin{eqnarray}
\label{pro}
P_{|\psi\rangle}(a_{1},a_{2},\cdots,a_{n}|\mathbf{x}_{1},\mathbf{x}_{2},\cdots,\mathbf{x}_{n})=\langle\psi|\prod_{i=1}^{n}E^{a_{i},\mathbf{x}_{i}}_{i}|\psi\rangle,
\end{eqnarray}
where $E^{a_{i},\mathbf{x}_{i}}_{i}$ are defined bellow.



We note that the kinematic postulates of standard quantum mechanics are as follows:

$\textbf{(1)}$ $|\psi\rangle$ is a normalized state vector in an arbitrary finite-dimensional Hilbert space of the composite physical system $\mathcal{H}=\mathcal{H}_{1}\otimes\ldots\otimes\mathcal{H}_{n}$.

$\textbf{(2)}$ A complete set of orthogonal projectors $\left\{E^{a_{i},\mathbf{x}_{i}}_{i}\right\}$ acting on the state vector $|\psi\rangle\in\mathcal{H}$ indicate the
measurements by the parties.

These operations satisfy $E^{a_{i},\mathbf{x}_{i}}_{i}E^{a'_{i},\mathbf{x}_{i}}_{i}=\delta_{a'_{i}, a_{i}} E^{a_{i},\mathbf{x}_{i}}_{i}$ and $\sum_{a_{i}}E^{a_{i},\mathbf{x}_{i}}_{i}=\openone_{i}$, where $\delta_{a'_{i}, a_{i}}$ is the Kronecker delta function and $\openone_{i}$ is the identity operation for the party $i$. Here we have used a simplified notation to present the projection operators with real elements $\openone_{1}\otimes\cdots\otimes E^{a_{i},\mathbf{x}_{i}}_{i}\otimes\cdots\otimes \openone_{n}=:E^{a_{i},\mathbf{x}_{i}}_{i},\hspace{.2cm}\forall i ,\hspace{.3cm}i=1,\cdots,n$. Thus, every sequence of the projection operators can be represented as
$\left(\openone_{1}\otimes\cdots\otimes E^{a_{i},\mathbf{x}_{i}}_{i}\otimes\cdots\otimes \openone_{j}\otimes\cdots\otimes \openone_{n}\right)\otimes\left(
\openone_{1}\otimes\cdots\otimes \openone_{i}\otimes\cdots\otimes E^{a_{j},\mathbf{x}_{j}}_{j}\otimes\cdots\otimes \openone_{n}\right)=
E^{a_{i},\mathbf{x}_{i}}_{i}E^{a_{j},\mathbf{x}_{j}}_{j}$.

$\textbf{(3)}$ Any pair of projection operators of different parties commute $\left[E^{a_{i},\mathbf{x}_{i}}_{i}, E^{a_{j},\mathbf{x}_{j}}_{j}\right]=0, \hspace{.1cm} i\neq j$.

$\vspace{.1cm}$


In the context of quantum mechanics, the isotropy/homogeneity symmetry transformations are represented by the unitary rotation/translation transformations on the state vectors and the projection operators such that the joint and marginal probability distributions (hence correlation functions) remain invariant. The unitary transformations of the $i$th party are represented by $U_{i}(R(\theta_{i},\mathbf{k}_{i}))=:U_{i}(\theta_{i},\mathbf{k}_{i})$, which are representations of the group of local rotations/translations determined by the direction $\mathbf{k}_{i}$ with a finite value (angle/distance) $\theta_{i}$. In fact, for simplicity and without risk of ambiguity, we imply a single notation to represent both rotation angle and translation distance. The local unitary operators $\left\{U_{i}(\theta_{i},\mathbf{k}_{i}),\hspace{.1cm}i=1,\cdots,n\right\}$ transform the set comprised of the initial state vector and the projection operators $(|\psi\rangle, \{E^{a_{i},\mathbf{x}_{i}}_{i},\hspace{.1cm}i=1,\cdots,n\})$ to the new set $(|\psi'\rangle, \{E^{a_{i},\mathbf{x}'_{i}}_{i},\hspace{.1cm}i=1,\cdots,n\})$ such that:
\begin{eqnarray}\label{rot}
&&|\psi\rangle\rightarrow |\psi'\rangle=\prod_{i=1}^{n}U_{i}(\theta_{i},\mathbf{k}_{i})|\psi\rangle,\nonumber\\
&&E^{a_{i},\mathbf{x}_{i}}_{i}\rightarrow E^{a_{i},\mathbf{x}'_{i}}_{i}=U_{i}(\theta_{i},\mathbf{k}_{i})\,E^{a_{i},\mathbf{x}_{i}}_{i}\,U_{i}^{\dag}(\theta_{i},\mathbf{k}_{i}),
\hspace{.3cm}\forall\hspace{.1cm} a_{i},\, \theta_{i},\, \mathbf{x}_{i},\, \mathbf{k}_{i},\hspace{.2cm}i=1,\cdots,n.
\end{eqnarray}
where, $U_{i}(\theta_{i},\mathbf{k}_{i})U_{i}^{\dag}(\theta_{i},\mathbf{k}_{i})=
U_{i}^{\dag}(\theta_{i},\mathbf{k}_{i})U_{i}(\theta_{i},\mathbf{k}_{i})=\openone, \forall\hspace{.1cm}i$. Here the measurement inputs $\mathbf{x}'_{i}$ is given by rotating/translating the measurement inputs $\mathbf{x}_{i}$ in the direction of $\mathbf{k}_{i}$ by the finite value $\theta_{i}$, $\mathbf{x}'_{i}
=R(\theta_{i},\mathbf{k}_{i})\mathbf{x}_{i}$.



In addition to postulates $(\mathbf{1}),(\mathbf{2})$ and $(\mathbf{3})$, we note that the isotropy and the homogeneity symmetries of the flat space are the fundamental principles which characterize quantum correlations. We can present the isotropy/homogeneity symmetry principle in standard quantum mechanics in the form of another postulate ($\mathbf{4}$) as follows:

$\textbf{(4)}$ The joint and the marginal probability distributions remain invariant under the isotropy/homogeneity symmetry transformations (\ref{rot}),
\begin{eqnarray}\label{pro1}
\langle\psi'|\prod_{i=1}^{n}E^{a_{i},\mathbf{x}'_{i}}_{i}|\psi'\rangle
=\langle\psi|\prod_{i=1}^{n}E^{a_{i},\mathbf{x}_{i}}_{i}|\psi\rangle,
\hspace{.3cm}\forall\hspace{.1cm} a_{i},\, \mathbf{x}_{i},\hspace{.2cm}i=1,\cdots,n.
\end{eqnarray}
An immediate consequence is the equality of the marginal probability distributions
$$\sum_{a_{i},\cdots,a_{j}}\langle\psi'|\prod_{i=1}^{n}E^{a_{i},\mathbf{x}'_{i}}_{i}|\psi'\rangle
=\sum_{a_{i},\cdots,a_{j}}\langle\psi|\prod_{i=1}^{n}E^{a_{i},\mathbf{x}_{i}}_{i}|\psi\rangle,$$
where the summation is taken on arbitrary subsets of the measurement outputs.

In fact, in the standard quantum mechanics equations ($\mathbf{3}$) (or postulate $\mathbf{4}$) are obviously manifested by  the following commutation relation:
\begin{eqnarray}\label{com}
\left[E^{a_{i},\mathbf{x}_{i}}_{i}, U_{j}(\theta_{j},\mathbf{k}_{j})\right]=\left[U_{i}(\theta_{i},\mathbf{k}_{i}), U_{j}(\theta_{j},\mathbf{k}_{j})\right]=0, \hspace{.3cm}\forall\hspace{.1cm} i\neq j, \hspace{.2cm}i,j=1,\cdots,n.
\end{eqnarray}
These relations can be proved by spectral decomposition of the unitary transformations as functions of the projection operators
$U_{j}(\theta_{j},\mathbf{k}_{j})=\sum_{a_{j}}
e^{ic(\theta_{j},\mathbf{k}_{j},a_{j})}E^{a_{j},\mathbf{k}_{j}}_{j}$, where the coefficients $e^{ic(\theta_{j},\mathbf{k}_{j},a_{j})}$ are the eigenvalues of $U_{j}$.



$\vspace{.1cm}$

\emph{The AQCs model}.---The isotropy/homogeneity principle plays an inherent role in every physical theory including standard quantum mechanics. Hence, it is natural to expect that even the AQCs model should respect these symmetries of the flat space. We now recall the postulates of the AQCs model (denoted shortly by ``$\tilde{Q}$") according to Ref. \cite{Alm} and examine the isotropy and homogeneity symmetries in this model. Here are the postulates

$(\textbf{1}')$ $|\psi\rangle$ is a normalized state vector in a Hilbert space of a composite physical system $\mathcal{H}=\mathcal{H}_{1}\otimes\ldots\otimes\mathcal{H}_{n}$.

$(\textbf{2}')$ A complete set of orthonormal projectors $\left\{E^{a_{i},\mathbf{x}_{i}}_{i}\right\}$ act on the state vector $|\psi\rangle$ and they indicate the
measurements performed by each party. These projection operators satisfy $E^{a_{i},\mathbf{x}_{i}}_{i}E^{a'_{i},\mathbf{x}_{i}}_{i}=\delta_{a'_{i}, a_{i}} E^{a_{i},\mathbf{x}_{i}}_{i}$ and $\sum_{a_{i}}E^{a_{i},\mathbf{x}_{i}}_{i}=\openone$.

$(\textbf{3}')$ Action of any arbitrary permutation of the projection operators on the state vector $|\psi\rangle$ are equal
\begin{eqnarray*}\label{}
E^{a_{1},\mathbf{x}_{1}}_{1}\cdots E^{a_{n},\mathbf{x}_{n}}_{n}|\psi\rangle
=\pi\left(E^{a_{1},\mathbf{x}_{1}}_{1}\cdots E^{a_{n},\mathbf{x}_{n}}_{n}\right)|\psi\rangle,\, \forall \, \pi.
\end{eqnarray*}
The permutation operator $\pi$ for the case of two parties with operators $A_{i}$ and $B_{j}$, respectively, is defined as $\pi(A_{i}B_{j})|\psi\rangle=A_{i}B_{j}|\psi\rangle=B_{j}A_{i}|\psi\rangle,\hspace{.1cm}\forall\hspace{.1cm} i,j=1,\cdots,n, \hspace{.2cm}i\neq j$, where $A_{i}$ and $B_{j}$ can be projection operators. However, in this article, it suffices to simply denote the permutation operators by a single notation $\pi$. The set of AQCs satisfies all of the kinematic postulates of quantum mechanics expect one; postulate ($\mathbf{3}$), which is about the commutativity of the projection operators of different parties, is replaced by postulate ($\mathbf{3}'$), which trades commutativity of permeation.

The set $\tilde{Q}$ is strictly larger than the set of quantum correlations and violates quantum upper bounds in both bipartite and multipartite systems \cite{Alm}. The remark that Ref. \cite{Alm} on the  $\tilde{Q}$ model does not explicitly specify whether postulate ($\mathbf{3}'$) holds for arbitrary set of measurement inputs $\mathbf{x}_{i},\hspace{.1cm}i=1,\cdots,n$, or not. In what follows, we explicitly prove this property as a consequence of the flat space symmetries, which shall enable us to show how the AQCs model reduces to quantum mechanics.



$\vspace{.1cm}$

\emph{The space isotropy/homogeneity symmetries generate infinite set of $\tilde{Q}$ models}.---As pointed out in Ref. \cite{Alm}, to give a complete information about the $\tilde{Q}$ model ``one should probe the action of non-commutative projection operators on states different from $|\psi\rangle$." However, we should note that in the $\tilde{Q}$ model, we only have accessed to the particular state vector  $|\psi\rangle$ and its associated probability distribution $P_{|\psi\rangle}(a_{1},a_{2},\cdots,a_{n}|\mathbf{x}_{1},\mathbf{x}_{2},\cdots,\mathbf{x}_{n})$. In order to give complete information of other probability distributions, we import the isotropy/homogenity symmetry transformations (\ref{rot}) in to the $\tilde{Q}$ scenario. We then impose these symmetry transformations on the AQCs model and request that the probability distributions remain invariant.

Similar to standard quantum mechanics, the local unitary operators $\{U_{i}(\theta_{i},\mathbf{k}_{i}), \hspace{.1cm}\forall\hspace{.1cm} i\}$ transform the set consisting of the initial state vector and the projection operators $\{|\psi\rangle, E^{a_{i},\mathbf{x}_{i}}_{i},\hspace{.1cm}\forall\hspace{.1cm} i\}$ to the new set $\{|\psi'\rangle, E^{a_{i},\mathbf{x}'_{i}}_{i},\hspace{.1cm}\forall\hspace{.1cm} i\}$, which are requested to satisfy postulates $(\mathbf{1}'), (\mathbf{2}')$, and $(\mathbf{3}')$. The symmetry transformations (\ref{rot}) in the $\tilde{Q}$ scheme are represented by
\begin{eqnarray}\label{rot1}
&&E^{a_{i},\mathbf{x}'_{i}}_{i}=U_{i}(\theta_{i},\mathbf{k}_{i})E^{a_{i},\mathbf{x}_{i}}_{i}U_{i}^{\dag}(\theta_{i},\mathbf{k}_{i}),\hspace{.1cm} \forall\hspace{.2cm}\theta_{i}, \hspace{.1cm}\mathbf{k}_{i},\hspace{.3cm}i=1,\cdots,n, \hspace{2cm} (i)\\\nonumber
&&|\psi'\rangle=\prod_{i=1}^{n}U_{i}(\theta_{i},\mathbf{k}_{i})|\psi\rangle
=\pi\left(\prod_{i=1}^{n}U_{i}(\theta_{i},\mathbf{k}_{i})\right)|\psi\rangle,\hspace{.1cm} \forall\hspace{.1cm} \pi \hspace{4cm} (ii)\nonumber\\\nonumber
&&E^{a_{1},\mathbf{x}'_{1}}_{1}\cdots E^{a_{n},\mathbf{x}'_{n}}_{n}|\psi'\rangle=\pi\left(E^{a_{1},\mathbf{x}'_{1}}_{1}\cdots E^{a_{n},\mathbf{x}'_{n}}_{n}\right)|\psi'\rangle,\hspace{.1cm} \forall\hspace{.1cm} \pi\hspace{4cm} (iii)\\\nonumber
&&\prod_{i\in I}U_{i}(\theta_{i},\mathbf{k}_{i})\prod_{j\in J}E^{a_{j},\mathbf{x}_{j}}_{j}|\psi\rangle
=\pi\left(\prod_{i\in I}U_{i}(\theta_{i},\mathbf{k}_{i})\prod_{j\in J}E^{a_{j},\mathbf{x}_{j}}_{j}
\right)|\psi\rangle,\hspace{.1cm} \forall\hspace{.1cm} \pi. \hspace{1cm}(iv)
\end{eqnarray}
In the $\tilde{Q}$ scenario, we note that the usual representation of the permutation operator $\pi$ \cite{Alm} is not extendable to combination of the projection operators and the unitary transformations \cite{App}. Although in what follows we do not require explicit representation of $\pi$, it is similar to what define in postulate ($\mathbf{3}'$), in which $A_{i}$ and $B_{j}$ are the projection operators or the unitary transformations. The permutation relations (\ref{rot1}-$ii$) and (\ref{rot1}-$iv$) represent counterparts of the commutation relations (\ref{com}) in standard quantum mechanics. In Eq. (\ref{rot1}-$iv$), the sets $I$ and $J$ are two arbitrary and disjoint ($I \cap J=\varnothing$) subsets of the parties $\big(I\subset\{1,\cdots,n\}, J\subset\{1,\cdots,n\}\big)$, where the number of elements of every one of the subsets ranges as $0\leqslant I, J\leqslant n$---see Ref. \cite{App} for explicit example. Besides, this equation holds in rotated/translated frame with transferring $\mathbf{x}_{i}\rightarrow\mathbf{x}'_{i},\,\forall\,i$, and
$|\psi\rangle\rightarrow|\psi'\rangle$. It is evident that Eqs. (\ref{rot1}-$ii$) and (\ref{rot1}-$iii$) are especial cases of Eq. (\ref{rot1}-$iv$), for ($I, J$)=($n, 0$) and ($I, J$)=($0, n$), respectively. Moreover, in the Eq. (\ref{rot1}-$ii$), the complete set of elements of Hilbert space $\mathcal{H}=\bigotimes_{i=1}^{n}\mathcal{H}_{i}$ can be generated by variation of the rotation/translation parameter $\theta_{i}$ and around direction $\mathbf{k}_{i}$ $i=1,\cdots,n$ which every state vector of the Hilbert space is defined as $\left\{|\Psi\rangle|\hspace{.1cm}|\Psi\rangle=\prod_{i=1}^{n}U_{i}(\theta_{i},\mathbf{k}_{i})|\psi\rangle,\hspace{.1cm}\forall\hspace{.1cm}\theta_{i}, \mathbf{k}_{i}, \hspace{.1cm} i=1,\cdots,n \right\}$.

Equations (\ref{rot1}) offer nobel ways to give complete information about the action of projection operators on states other than $|\psi\rangle$. In particular, each of the Eqs. (\ref{rot1}-$ii$) and (\ref{rot1}-$iv$) are sufficient to achieve this goal and reduce the set $\tilde{Q}$ to standard quantum mechanics; see Appendix \ref{E}. Nevertheless, as a consequence of the symmetries of the flat space (postulate ($\mathbf{4}'$) bellow) and Eq. (\ref{rot1}-$iii$), we show that postulate ($\mathbf{3}'$) holds for arbitrary sets of measurement inputs $\mathbf{x}_{i},\, \forall\, i$, (bellow (\ref{pro81})). This relation enables us to reduce the set $\tilde{Q}$ to standard quantum mechanics, and additionally, Eqs. (\ref{rot1}-$ii$) and (\ref{rot1}-$iv$) shall be natural outcomes of the postulate ($\mathbf{3}'$); see Appendix \ref{D}.





We now modify postulate ($\mathbf{4}$) in the $\tilde{Q}$ scenario as follows:


$(\textbf{4}')$ The joint and the marginal probability distributions remain invariant under arbitrary rotation/translation unitary transformations $U_{i}(\theta_{i},\mathbf{k}_{i}), i=1,\cdots,n$ on the state vector $|\psi\rangle$ and the projection operators $E^{a_{i},\mathbf{x}_{i}}_{i}$ (\ref{rot1}):
\begin{eqnarray}\label{pro2}
&&\langle\psi|\prod_{i=1}^{n}E^{a_{i},\mathbf{x}_{i}}_{i}|\psi\rangle=\langle\psi|\pi\left(\prod_{j=1}^{n}E^{a_{j},\mathbf{x}_{j}}_{j}\right)|\psi\rangle
=\langle\psi'|\prod_{j=1}^{n}E^{a_{j},\mathbf{x}'_{j}}_{j}|\psi'\rangle
=\langle\psi'|\pi\left(\prod_{j=1}^{n}E^{a_{j},\mathbf{x}'_{j}}_{j}\right)|\psi'\rangle\nonumber
\\
&&=\langle\psi|\prod_{i=1}^{n}U_{i}^{\dagger}(\theta_{i},\mathbf{k}_{i})
\prod_{j=1}^{n}\left(U_{j}(\theta_{j},\mathbf{k}_{j})E^{a_{j},\mathbf{x}_{j}}_{j}U_{j}^{\dag}(\theta_{j},\mathbf{k}_{j})\right)
\prod_{k=1}^{n}U_{k}(\theta_{k},\mathbf{k}_{k})|\psi\rangle\\\nonumber
&&=\langle\psi|\pi_{L}\left(\prod_{i=1}^{n}U_{i}^{\dagger}(\theta_{i},\mathbf{k}_{i})\right)
\pi_{C}\left(\prod_{j=1}^{n}U_{j}(\theta_{j},\mathbf{k}_{j})E^{a_{j},\mathbf{x}_{j}}_{j}U_{j}^{\dag}(\theta_{j},\mathbf{k}_{j})\right)
\pi_{R}\left(\prod_{k=1}^{n}U_{k}(\theta_{k},\mathbf{k}_{k})\right)|\psi\rangle,\hspace{.1cm} \forall\hspace{.1cm} \pi.
\end{eqnarray}
In the last line, we have added indexes $L,R$ for the permutation operators $\pi$ to distinguish permutations of the unitary operators acting on $\langle\psi|$ and $|\psi\rangle$, respectively. Moreover, we used the index $C$ in the middle ($\pi_{C}$) to represent permutation of $E^{a_{j},\mathbf{x}'_{j}}_{j}\, \forall \,j$ on the state $|\psi'\rangle$. Note that disjoint operators $E_{i}$ (or $U_{i}$, $U_{i}^{\dag}$) and $E_{j}$ (or $U_{j}$, $U_{j}^{\dag}$), $\forall\, i\neq j$, can be permuted. But, the order of $U_{i}$, $E_{i}$ and $U_{i}^{\dag}$, $\forall\,i$ should be saved; in particular, $U_{i}$ ($U_{i}^{\dag}$) should always appear on the left (right) side of $E_{i}$.

In addition, similar to  what stated after postulate ($\mathbf{4}$), the marginal probability distributions are equal,
\begin{eqnarray*}\label{}
\sum_{a_{i},\cdots,a_{j}}\langle\psi|\prod_{i=1}^{n}E^{a_{i},\mathbf{x}_{i}}_{i}|\psi\rangle
=\sum_{a_{i},\cdots,a_{j}}\langle\psi'|\prod_{j=1}^{n}E^{a_{j},\mathbf{x}'_{j}}_{j}|\psi'\rangle,
\end{eqnarray*}
where the summation is on arbitrary subsets of the measurement outputs.



There is an equivalent relation between the invariance of the probability distributions (\ref{pro2}) and the symmetry transformations (\ref{rot1}), which can be represented in the following theorem:

\textbf{Theorem $\mathbf{1}$}: In the original $\tilde{Q}$ model, the probability distributions remain invariant $\langle\psi|\prod_{i=1}^{n}E^{a_{i},\mathbf{x}_{i}}_{i}$ $|\psi\rangle
=\langle\psi'|\prod_{j=1}^{n}E^{a_{j},\mathbf{x}'_{j}}_{j}|\psi'\rangle$ [Eq. (\ref{pro2})] if and only if Eqs. (\ref{rot1}) hold.

\emph{Proof}: See Appendix \ref{D}. \hfill $\blacksquare$

In the following, to demonstrate our goal it suffices to prove the sufficient condition of this theorem and show that postulate ($\mathbf{3}'$) holds for arbitrary set of measurement inputs $\mathbf{x}_{i},\, \forall\, i$. This shall help us prove the theorem for general cases.







We now take a specific subset of the unitary transformations (\ref{rot1}) where one of the parties, say, the first $1$, undergoes a local rotation/translation transformation $U_{1}(\theta_{1},\mathbf{k}_{1}), \hspace{.1cm} \theta_{1}\neq0$, and other subsystems remain at rest, $U_{i}(\theta_{i},\mathbf{k}_{i})=\openone,\hspace{.1cm} \theta_{i}=0,\hspace{.1cm}i=2,\cdots,n$, and prove the following theorem:

\textbf{Theorem $\mathbf{2}$}: If Eq. (\ref{rot1}-$iv$) holds, then: (\emph{$\mathbf{O_{1}}$}) the probability distributions remain invariant, (\emph{$\mathbf{O_{2}}$}) some other can also sets of $\tilde{Q}$ models can be generated, and (\emph{$\mathbf{O_{3}}$}) postulate ($\mathbf{3}'$) holds for arbitrary sets of measurement inputs $\mathbf{x}_{i},\, \forall\, i$.

\emph{Proof}: (\emph{$\mathbf{O_{1}}$}): The probability distribution (\ref{pro2}) is given by
\begin{eqnarray}\label{pro3}
&&\langle\psi|\prod_{i=1}^{n}E^{a_{i},\mathbf{x}_{i}}_{i}|\psi\rangle=\langle\psi'|\prod_{i=1}^{n}E^{a_{i},\mathbf{x'}_{i}}_{i}|\psi'\rangle
\\\nonumber
&&\hspace{2.9cm}=\langle\psi|E^{a_{1},\mathbf{x}_{1}}_{1}U_{1}^{\dag}(\theta_{1},\mathbf{k}_{1})
\left(\prod_{j=2}^{n}E^{a_{j},\mathbf{x}_{j}}_{j}\right)U_{1}(\theta_{1},\mathbf{k}_{1})|\psi\rangle,
\hspace{.3cm}\forall\hspace{.1cm}\theta_{1}, \hspace{.1cm}\mathbf{k}_{1}.
\end{eqnarray}
According to Eq. (\ref{rot1}-$iv$), if arbitrary permutations of $U_{1}(\theta_{1},\mathbf{k}_{1})$ and $E^{a_{j},\mathbf{x}_{j}}_{j}$ ($j=2,\cdots,n$) acting on the state $|\psi\rangle$ are assumed equal,
\begin{eqnarray}\label{pro6}
\prod_{j=2}^{n}E^{a_{j},\mathbf{x}_{j}}_{j}U_{1}(\theta_{1},\mathbf{k}_{1})|\psi\rangle
=\pi\left(\prod_{j=2}^{n}E^{a_{j},\mathbf{x}_{j}}_{j}U_{1}(\theta_{1},\mathbf{k}_{1})\right)|\psi\rangle,
\end{eqnarray}
the first equality in Eq. (\ref{pro3}) holds, if $U_{1}$ moves to the left side of $E^{a_{j},\mathbf{x}_{j}}_{j},\, j=2,\cdots,n$.


(\emph{$\mathbf{O_{2}}$}): Similar to postulate ($\mathbf{4}$), in Eq. (\ref{pro6}), we use the spectral decomposition of $U_{1}$, $U_{1}(\theta_{1},\mathbf{k}_{1})=\sum_{a_{1}}e^{ic(\theta_{1},\mathbf{k}_{1},a_{1})} E^{a_{1},\mathbf{k}_{1}}_{1}$,
which yields

\begin{align}\label{pro7}
\sum_{a_{1}}e^{ic(\theta_{1},\mathbf{k}_{1},a_{1})}
\prod_{j=2}^{n}E^{a_{j},\mathbf{x}_{j}}_{j}
E^{a_{1},\mathbf{k}_{1}}_{1}|\psi\rangle
=\sum_{a_{1}}e^{ic(\theta_{1},\mathbf{k}_{1},a_{1})}\pi\left(
\left[\prod_{j=2}^{n}E^{a_{j},\mathbf{x}_{j}}_{j}\right]E^{a_{1},\mathbf{k}_{1}}_{1}\right)|\psi\rangle.
\end{align}
As pointed out in postulate ($\mathbf{2}'$), because the projection operators $\{E^{a_{1},\mathbf{k}_{1}}_{1}\}$ represent a complete set of orthonormal basis, the permutation relations hold for each of the projection operators $E^{a_{1},\mathbf{k}_{1}}_{1} \hspace{.1cm} (a_{1}=1,\cdots,d_{1})$ and $E^{a_{j},\mathbf{x}_{j}}_{j}, \hspace{.2cm} j\neq1$,
\begin{eqnarray}\label{pro8}
\left[\prod_{j=2}^{n}E^{a_{j},\mathbf{x}_{j}}_{j}\right]E^{a_{1},\mathbf{k}_{1}}_{1}|\psi\rangle
=\pi\left(\left[\prod_{j=2}^{n}E^{a_{j},\mathbf{x}_{j}}_{j}\right]E^{a_{1},\mathbf{k}_{1}}_{1}\right)|\psi\rangle.
\end{eqnarray}
Therefore, we generate a set of $\tilde{Q}$ models in which the first party input $\mathbf{x}_{1}$ is replaced by $\mathbf{k}_{1}$ in postulate ($\mathbf{3}'$).

(\emph{$\mathbf{O_{3}}$}): We can repeat a similar scenario to generate another set of $\tilde{Q}$ models for the new set of measurement inputs $\left\{\mathbf{k}_{1}, \mathbf{x}_{j}, \hspace{.2cm} j=2,\cdots,n\right\}$. For example, we apply a local rotation/translation transformation $U_{2}(\theta_{2},\mathbf{k}_{2}), \hspace{.1cm} \theta_{2}\neq0$ on the second subsystem, and the other subsystems remain at rest, $U_{i}(\theta_{i},\mathbf{x}_{i})=\openone,\hspace{.1cm} \theta_{i}=0,\hspace{.1cm}i=1,3,4,\cdots,n$, and give another set of permutation relations as
\begin{eqnarray*}\label{}
\left[\prod_{j=3}^{n}E^{a_{j},\mathbf{x}_{j}}_{j}\right]E^{a_{2},\mathbf{k}_{2}}_{2}E^{a_{1},\mathbf{k}_{1}}_{1}|\psi\rangle
=\pi\left(\left[\prod_{j=3}^{n}E^{a_{j},\mathbf{x}_{j}}_{j}\right]E^{a_{2},\mathbf{k}_{2}}_{2}E^{a_{1},\mathbf{k}_{1}}_{1}\right)|\psi\rangle.
\end{eqnarray*}
We can apply a similar procedure to the other subsystems and generate other sets of the permutation relations. This way, we eventually obtain
\begin{eqnarray}\label{pro81}
\prod_{j=1}^{n}E^{a_{j},\mathbf{k}_{j}}_{j}|\psi\rangle
=\pi\left(\prod_{j=1}^{n}E^{a_{j},\mathbf{k}_{j}}_{j}\right)|\psi\rangle, \hspace{.2cm} \forall \hspace{.1cm}a_{j},\, \mathbf{k}_{j}.
\end{eqnarray}
Therefore, we have proven that postulate (\textbf{$\mathbf{3}'$}) holds for \emph{any} set of measurement inputs. \hfill $\blacksquare$



$\vspace{.1cm}$

\emph{The set of $\tilde{Q}$ models violate the space isotropy/homogeneity principle}.---
In the previous section, we extended postulate ($\mathbf{3}'$) for arbitrary input settings $\mathbf{x}'_{i},\hspace{.1cm}\forall\, i$,
\begin{eqnarray}\label{pro9}
E^{a_{1},\mathbf{x}'_{1}}_{1}\cdots E^{a_{n},\mathbf{x}'_{n}}_{n}|\psi\rangle
=\pi\left(E^{a_{1},\mathbf{x}'_{1}}_{1}\cdots E^{a_{n},\mathbf{x}'_{n}}_{n}\right)|\psi\rangle, \hspace{.2cm} \forall \hspace{.1cm}a_{i},\,\mathbf{x}'_{i},\hspace{.1cm} i=1,\cdots,n.
\end{eqnarray}

on the other hand, due to Eq. (\ref{rot1}-$iii$), permutation of the projection operators also hold for the rotated/translated measurement inputs and state vector $(\{\mathbf{x}_{i}\}, \hspace{.1cm}|\psi\rangle)\rightarrow (\{\mathbf{x}'_{i}\}, \hspace{.1cm} |\psi'\rangle), \hspace{.2cm}i=1,\cdots,n$,
\begin{eqnarray}\label{pro10}
E^{a_{1},\mathbf{x}'_{1}}_{1}\cdots E^{a_{n},\mathbf{x}'_{n}}_{n}|\psi'\rangle=\pi\left(E^{a_{1},\mathbf{x}'_{1}}_{1}\cdots E^{a_{n},\mathbf{x}'_{n}}_{n}\right)|\psi'\rangle,\hspace{.3cm}\forall\hspace{.2cm}a_{i}, \hspace{.1cm}\mathbf{k}_{i}.
\end{eqnarray}

Now we apply Eq. (\ref{pro9}) to this relation and substitute the inputs $\mathbf{x}'_{i}$ with the inputs $\mathbf{x}_{i}$, whereby
\begin{eqnarray*}\label{}
E^{a_{1},\mathbf{x}_{1}}_{1}\cdots E^{a_{n},\mathbf{x}_{n}}_{n}|\psi'\rangle
=\pi\left(E^{a_{1},\mathbf{x}_{1}}_{1}\cdots E^{a_{n},\mathbf{x}_{n}}_{n}\right)|\psi'\rangle.
\end{eqnarray*}
This shows that the state $|\psi'\rangle$ plays the same role as $|\psi\rangle$ in postulate ($\mathbf{3}'$). We can cover whole Hilbert space $\mathcal{H}=\mathcal{H}_{1}\otimes\ldots\otimes\mathcal{H}_{n}$ by changing the rotation/translation parameters $\theta_{i}$ and directions $\mathbf{k}_{i},\hspace{.1cm} i=1,\cdots,n$. Therefore, postulate ($\mathbf{3}'$) holds for every arbirtaty vector of the Hilbert space, $\forall\, |\Psi\rangle\in\mathcal{H}$,
\begin{eqnarray*}\label{pro11}
\prod_{i=1}^{n}E^{a_{i},\mathbf{x}_{i}}_{i}|\Psi\rangle
=\pi\left(\prod_{i=1}^{n}E^{a_{i},\mathbf{x}_{i}}_{i}\right) |\Psi\rangle, \hspace{.2cm}
\left\{|\Psi\rangle||\Psi\rangle=\prod_{i=1}^{n}U_{i}(\theta_{i},
\mathbf{k}_{i})|\psi\rangle,\hspace{.1cm}\forall\hspace{.1cm} \theta_{i}, \mathbf{k}_{i}, \hspace{.1cm} i=1,\cdots,n \right\}.
\end{eqnarray*}

For any pair of projection operators $E^{a_{i},\mathbf{x}_{i}}_{i}$ and $E^{a_{j},\mathbf{x}_{j}}_{j},\,i\neq j$, the permutation relations can be rewritten as the commutator of these projections on any vector of the Hilbert space $\mathcal{H}$,
\begin{eqnarray}\label{pro110}
\left[E^{a_{i},\mathbf{x}_{i}}_{i},E^{a_{j},\mathbf{x}_{j}}_{j}\right]|\Psi\rangle=0,
\hspace{.1cm}\forall\hspace{.1cm}|\Psi\rangle\in\mathcal{H},\hspace{.1cm}\forall\hspace{.1cm} \mathbf{x}_{i},\hspace{.1cm} \mathbf{x}_{j}, \hspace{.2cm} i\neq j,\hspace{.2cm}i,j=1,\cdots,n,
\end{eqnarray}
which is equivalent to the commution relation $\left[E^{a_{i},\mathbf{x}_{i}}_{i},E^{a_{j},\mathbf{x}_{j}}_{j}\right]=0$.
Hence postulate ($\mathbf{3}'$) in the $\tilde{Q}$ model reduces to postulate ($\mathbf{3}$) in standard quantum mechanics.

$\vspace{.1cm}$

\emph{The isotropy/homogenuty principle enables to characterize sets of multipartite quantum correlations from the multipartite versions of the AQCs models}.---The existing information-theoretic principles \cite{Dam,Bra2,NC,ML,ML1,IC,IC2,IC3,MBL,LO,Alm1,Alm3} supply fundamental limitations on post-quantum models \cite{PR1,BCP,JM,BP1,BP2,Mas,Alm}, which aim to bring their predictions closer to quantum-mechanical predictions. However, these principles only provide partial answers to determination of bipartite correlation functions. In multipartite systems, this issue becomes even more challenging. Specifically, we can divide a multipartite system to subsystems in which the correlations of any bipartite subsystem admits a classical model, whereas the multipartite correlations are stronger than multipartite quantum correlations \cite{NIP,NIP1}. This scenario, which is referred to as time-ordered bi-local set \cite{NIP,NIP1}, respects all bipartite information-theoretic principles \cite{Dam,Bra2,NC,ML,ML1,IC,IC2,IC3,MBL,LO,Alm1,Alm3}. More importantly, this scenario implies that yet a fundamentally multipartite principle is needed for a more complete understanding of quantum predictions.

The AQCs model has partially addressed this problem by proposing a tripartite version of the Bell-type experiment \cite{Alm}.
Specifically, as an example, a tripartite Bell-type inequality \cite{Mun} with two-input and two-output setting for each party  ($n=3,m_{i}=2,d_{i}=2,\hspace{.1cm}i=1,2,3$) was considered and it was shown that the AQCs model predicts the value $10.14955$. However, this contradicts the value  $10.00217$, which is the quantum-mechanical prediction for the maximum value of Bell violation, by an amount equal to $0.147$ \cite{Alm}. Note that the tripartite AQCs model satisfies all information-theoretic principles \cite{Dam,Bra2,NC,ML,ML1,IC,IC2,IC3,MBL,LO,Alm1,Alm3}. This indicates that the bipartite information-theoretic principles fail to conclusively distinguish these types of post-quantum models from quantum mechanics. In addition, as pointed out in Ref. \cite{Alm}, there is an extension of the time-ordered bi-local set to the time-ordered bi-\emph{quantum} set where the probability distribution of the bi-partite subsystems behave quantum mechanically rather than classically. It predicts that the time-ordered bi-quantum set and standard quantum mechanics are indistinguishable in the level of bipartite systems, meanwhile there exist tripartite post-quantum models in the intersection of the time-ordered bi-quantum set and AQCs model. Thus, as it was indicated in Refs. \cite{NIP,NIP1}, there is still a need for true principle to remove this ambiguity, and that principle should have a genuinely multipartite feature.

Our approach in this paper, invoking the isotropy/homogenuty symmetry as the natural fundamental principle of any physical model, includes \emph{multipartite} AQCs models, and is inherently multipartite. Due to the symmetry transformations (\ref{rot1}) and postulate ($\mathbf{4}'$), the number of parties ($n$), the dimension of the input space ($m_{i},\hspace{.1cm}\forall\hspace{.1cm}i$), and the dimension of the output space ($d_{i},\hspace{.1cm}\forall\hspace{.1cm}i$) do not play any key role in the derivation of the commutation relation (\ref{pro110}). This property indicates that our scenario rules out every set of AQCs model or every extension of such models which belong to the time-ordered bi-local/bi-quantum sets.

$\vspace{.1cm}$

\emph{Discussion and outlook}.---We revisiting quantum mechanics form different perspectives is an important line of research in contemporary physics. In this paper, we have particulary highlighted a connection between degree of nonlocality in quantum mechanics and the isotopy/homogenuty symmetry of the flat space.
Specifically, we have proved the essential role of these symmetries of the space in reducing one post-quantum model (the almost quantum correlations) to standard quantum mechanics. This AQCs model satisfies all but one of the kinematic axioms of standard quantum correlations; nevertheless, it strictly allows stronger-than-quantum correlations. We have applied the quantum-mechanical representations of the isotropy/homogeneity symmetry group on the set of AQCs model and requested that the conditional probability distributions (the Born rule) remain invariant. We have proved that this condition is already sufficient to reduce the AQCs models to standard quantum mechanics. Moreover, we have highlighted that since our framework is by construction multipartite, and therein the number of parties does not play any role, our framework goes beyond tripartite version of AQCs model and rule them out.



Several natural and related questions can come up, which be hope our approach can help or shed some light on:

($i$) Is the connection between the nonlocality and the flat-space symmetries restricted only to AQCs model, or dose there exist a fundamental relation between these two concepts even beyond this particular model?

($ii$) The Tsirelson theorem indicates that the Heisenberg uncertainty principle is sufficient to violate the CHSH inequality at the maximum value $2\sqrt{2}$ \cite{Tsi}. Can the uncertainty principle itself be considered as a natural consequence of the symmetries of the flat space?

($iii$) Wether or how is it possible to extend our results to other types of post-quantum models \cite{Non1,Non2,Non3,Non4,Sv}?

($iv$) It is also interesting to investigate the impact of the isotropy/homogeneity symmetry on the generalized probabilistic theories \cite{GPT,GPT1,GPT2} and obtain a device-independent definition for space symmetries. It can enable to extend our results to more general cases. This is the subject of another study \cite{NLB}.



\emph{Acknowledgements}.---We  would like to thank A. T. Rezakhani for interesting discussions and valuable comments.

\newpage
\vspace{.7cm}

\centerline{\large \bf {Appendix}}

\vspace{.2cm}

In the appendixes, we formally state and prove the results mentioned in the main text. It is organized as follows. In App. \ref{A}, we briefly review the mathematical background of the isotropy/homogeneity symmetry transformations and their representations in quantum mechanics. In App. \ref{B}, we consider the isotropy/homogeneity symmetry properties in the Bell-type experiments and discusses how the rotation/translation symmetry are realized in these experiments.  In App. \ref{C}, we present explicit representation of the permutation operators for combination of projections operators and unitary transformations. In App. \ref{D}, we prove that Theorem $1$ of the main text.
Finally, in App. \ref{E}, we show that  the permutation relations (\ref{rot1}-$ii$) and (\ref{rot1}-$iv$) are sufficient to reduce the  $\tilde{Q}$ model to standard quantum mechanics.







\subsection{Mathematical background: Isotropy/homogeneity symmetry transformations in standard quantum mechanics}
\label{A}

In standard quantum mechanics, the isotropy and the homogeneity symmetry group transformations are represented by the unitary transformations $U(\theta,\mathbf{k})\in\left\{\mathbf{U}\right\}$, which include and represent local rotation/translation about direction $\mathbf{k}$ by a finite angle/distance $\theta$. Wether a $U$ is rotation or translation is clear from the context. The isotropy/homogeneity symmetry transformations satisfy the following group properties:

(\emph{i}) Closed under product of group elements. The product of any two elements $U(\theta_{i},\mathbf{k}_{i}), U(\theta_{j},\mathbf{k}_{j})\in\left\{\mathbf{U}\right\}$ is another element of the symmetry transformations, $U(\theta_{i},\mathbf{k}_{i}) U(\theta_{j},\mathbf{k}_{j})\in\left\{\mathbf{U}\right\}$.

(\emph{ii}) Existence of an identity transformation. There exists a unique element $\openone\in\left\{\mathbf{U}\right\}$ such that any element $U(\theta_{i},\mathbf{k}_{i})\in\left\{\mathbf{U}\right\}$ satisfies $U(\theta_{i},\mathbf{k}_{i}) \openone=\openone U(\theta_{i},\mathbf{k}_{i})=U(\theta_{i},\mathbf{k}_{i})$.

(\emph{iii}) Existence of inverse transformations. For any element $U(\theta_{i},\mathbf{k}_{i})\in\left\{\mathbf{U}\right\}$, there exists a unique element $U^{-1}(\theta_{i},\mathbf{k}_{i})\in\left\{\mathbf{U}\right\}$ such that $U(\theta_{i},\mathbf{k}_{i})U^{-1}(\theta_{i},\mathbf{k}_{i})=U^{-1}(\theta_{i},\mathbf{k}_{i})U(\theta_{i},\mathbf{k}_{i})=\openone$.

(\emph{iv}) Associativity law. Any three elements of the symmetry transformations $U(\theta_{i},\mathbf{k}_{i}), U(\theta_{j},\mathbf{k}_{j}), $ 
$U(\theta_{k},\mathbf{k}_{k})\in\left\{\mathbf{U}\right\}$ satisfy the following associativity condition:
$$\left[U(\theta_{i},\mathbf{k}_{i})U(\theta_{j},\mathbf{k}_{j})\right]U(\theta_{k},\mathbf{k}_{k})
=U(\theta_{i},\mathbf{k}_{i})\left[U(\theta_{j},\mathbf{k}_{j})U(\theta_{k},\mathbf{k}_{k})\right].$$


In quantum mechanics, the explicit forms of the rotation/translation operators are, respectively, as follows:
\begin{eqnarray*}\label{Rot55}
&&U_{\mathrm{rotation}}(\theta,\mathbf{k})=\exp\left(\frac{-i\mathbf{J}\cdot\mathbf{k}\theta}{\hbar}\right),\\\nonumber
&&U_{\mathrm{translation}}(\theta,\mathbf{k})=\exp\left(\frac{-i\mathbf{P}\cdot\mathbf{k}\theta}{\hbar}\right),
\end{eqnarray*}
where $\mathbf{J}$ and $\mathbf{P}$ are, respectively, the representations of the angular momentum and the liner momentum in three dimensions.

As pointed out in the Eqs. (\ref{rot}) and (\ref{pro1}) of the main text, the probability distributions should remain invariant under arbitrary rotation/translation unitary transformations.

\subsection{Isotropy/homogeneity symmetry transformations in the Bell-type experiments}
\label{B}

The Bell-types inequalities have so far been realized in diverse platforms and experimental settings, such as optical photons \cite{Sha,Ant,Pan}, atomic systems \cite{Han,Win}, and superconducting circuits \cite{Sup}. Due to crucial role of the isotropy and homogeneity principles in our scenario, we briefly review realization of the isotropy/homogeneity symmetry transformations in experimental setting.

In realization of the CHSH inequality by entangled photons \cite{Sha,Ant,Pan,Ph,Ph1}, the qubits are represented by photon polarization state and the external devices are optical devices such as quarter-wave/half-wave plates which control photon polarization direction. In another type of realization of the CHSH inequality in experiment, the qubits are represented by two-level trapped ions in the ground and excited states \cite{Han,Win,NV,NV1,NV2,NV3,NV4}, for example electronic spin in the center of a nitrogen vacancy \cite{Han} or a single rubidium-$87$ atom in an optical dipole trap \cite{Win,NV}. In these experiments, the external fields, such as microwave pulses \cite{Han} or the polarization of lasers \cite{Win,NV}, control and rotate the state of the atomic systems. Similar to schemes \cite{Han,Win}, in presentation of qubit with superconducting circuit, the qubit is rotated and controlled by external microwave pulses \cite{Sup1,Sup2}.

In all of experimental test of the Bell-type inequalities \cite{Sha,Ant,Pan,Han,Win,Sup}, there is one-to-one relation between state vectors (qubits) in quantum mechanics formalism and physical objects in experiment (polarization of photon, atomic state, superconducting circuit). Similar one-to-one relation holds between unitary transformations (measurement operators) and controlling devices (measurement devices) such as optical devices, lasers, microwave pulses (single-photon detector, spin-dependent fluorescence, gated microwave tone). Then, it is logical that the isotropy/homogeneity symmetry transformations (\ref{rot}) and the postulate ($\mathbf{4}$) in quantum physics should have counterpart in experiments. These transformations are realized by rotating/translating physical systems (qubits) by external devices (optical devices, lasers, microwave pulses) and rotating/translating experimental apparatus (sources, channels, or other measurement settings of the lab) by experimenters.




\subsection{Presentation of permutation operators}
\label{C}


The AQCs model satisfies all of the kinematic postulates of quantum mechanics expect the commutator of projection operators is replaced by any permutation of these operators. There are different representations of the permutation group $S_{n}$. For example, one can represent a permutation of the symbols $x_{1},  x_{2}, \cdots,  x_{n}$ by the following notation \cite{Cau}:
\begin{eqnarray}
\label{Per}
\pi=\left(
      \begin{array}{cccc}
        x_{1} &  x_{2} & \cdots &  x_{n} \\
         \pi(x_{1}) & \pi(x_{2}) & \cdots & \pi(x_{n}) \\
      \end{array}
    \right).
\end{eqnarray}

As indicated in Ref. \cite{Alm}, the permutation operators $\pi$ act on strings of $E_{i}$ in the natural way as $\pi\left(E^{a_{1},\mathbf{x}_{1}}_{1}\cdots E^{a_{n},\mathbf{x}_{n}}_{n}\right)|\psi\rangle
=E^{a_{\pi(1)},\mathbf{x}_{\pi(1)}}_{\pi(1)}\cdots E^{a_{\pi(n)},\mathbf{x}_{\pi(n)}}_{\pi(n)}|\psi\rangle$.

Nevertheless, when a permutation of a combination of projection and unitary transformations is needed this notation does not work. In fact, in this paper we do not need an explicit presentation of the permutation operations; what we need is permutation of disjoint subsets of unitary transformation ($I$) and projection operators ($J$) is defined bellow:
\begin{eqnarray}\label{per1}
&&\prod_{i\in I}U_{i}(\theta_{i},\mathbf{k}_{i})\prod_{j\in J}E^{a_{j},\mathbf{x}_{j}}_{j}
=\pi\left(\prod_{i\in I}U_{i}(\theta_{i},\mathbf{k}_{i})\prod_{j\in J}E^{a_{j},\mathbf{x}_{j}}_{j}
\right)|\psi\rangle\nonumber\\
&&\hspace{1cm}=\prod_{i_{1}\in I_{1}}U_{i_{i}}(\theta_{i_{1}},\mathbf{k}_{i_{1}})\prod_{j_{1}\in J_{1}}E^{a_{j_{1}},\mathbf{x}_{j_{1}}}_{j_{1}}
\prod_{i_{2}\in I_{2}}U_{i_{2}}(\theta_{i_{2}},\mathbf{k}_{i_{2}})\prod_{j_{2}\in J_{2}}E^{a_{j_{2}},\mathbf{x}_{j_{2}}}_{j_{2}}|\psi\rangle\nonumber
\\\nonumber
\\\nonumber
&&\hspace{1cm}=\hspace{5cm} \cdots\nonumber\\\nonumber\\
&&\hspace{1cm}=\prod_{i_{1}\in I_{1}}U_{i_{i}}(\theta_{i_{1}},\mathbf{k}_{i_{1}})\prod_{j_{1}\in J_{1}}E^{a_{j_{1}},\mathbf{x}_{j_{1}}}_{j_{1}}
\cdots\prod_{i_{k}\in I_{k}}U_{i_{k}}(\theta_{i_{k}},\mathbf{k}_{i_{k}})\prod_{j_{k}\in J_{k}}E^{a_{j_{k}},\mathbf{x}_{j_{k}}}_{j_{k}}
|\psi\rangle\\\nonumber\\\nonumber\\\nonumber
&&\hspace{1cm}=\hspace{5cm} \cdots, \nonumber
\end{eqnarray}
where $I, \, I_{t},\, (t=1,\cdots,k)$ and $J, \, J_{t},\, (t=1,\cdots,k)$ are disjoint and arbitrary sunsets of set of $\{1,\cdots,n\}$ such that
\begin{eqnarray*}\label{}
&&I\subset\{1,\cdots,n\}, J\subset\{1,\cdots,n\},\hspace{.1cm}I\cap J=\varnothing\nonumber\\
&& I_{1}\cup I_{2}=I,\hspace{.2cm} I_{1}\cap I_{2}=\varnothing,
\hspace{.4cm} J_{1}\cup J_{2}=J,\hspace{.1cm} J_{1}\cap J_{2}=\varnothing\nonumber\\\nonumber\\\nonumber
&&\hspace{1cm}\hspace{5cm} \cdots\nonumber\\\nonumber\\\nonumber
&&\bigcup_{t}^{k}I_{t}=I,\hspace{.1cm}I_{t}\cap I_{s}=\varnothing,\hspace{.3cm} \bigcup_{t}^{k} J_{t}=J,
\hspace{.1cm} J_{t}\cap J_{s}=\varnothing,\hspace{.2cm}  t\neq s=1,\cdots, k.\nonumber\\\nonumber\\\nonumber
&&\hspace{1cm}\hspace{5cm} \cdots
\end{eqnarray*}

This sequence can be naturally continued. The above permutation relations imply that any disjoint subsets of the unitary transformations and the projection operators can be permuted on the state $|\psi\rangle$.

\subsection{Proof of Theorem $1$}
\label{D}

As pointed out in the main text, Theorem $2$ is sufficient to reduce the $\tilde{Q}$ model to standard quantum mechanics. Here, we prove Theorem $1$, which is a more complete version of Theorem $2$, in order to show the crucial impact of the flat space symmetries on quantum mechanics.

Before doing so, some points are in order:

1- The sufficient condition of Theorem $1$ is already enough to show that the set $\tilde{Q}$ reduces to standard quantum mechanics. However, for the sake of completeness we also prove that this condition is necessary as well.

2- Note that each of the arbitrary permutation relations (\ref{rot1}-$iii$) and (\ref{rot1}-$iv$) are sufficient to reduce the $\tilde{Q}$ model to standard quantum mechanics (see Sec. \ref{E}). Here, however, our proof does not use these relations.

3- Equations (\ref{rot1}-$iii$) and (\ref{rot1}-$iv$) can be derived as outcomes of the invariance of the permutation relations of the projection operators under the flat space symmetry transformations.

We first recall Theorem 1 and then prove it in three steps, the proof theorem $1$ is organized in three steps. First, we prove the theorem for specific a subset of the unitary transformations; next, we derive the permutation relations (\ref{rot1}-$ii$) and (\ref{rot1}-$iv$); and finally, we use these steps to prove the theorem in the general case.

$\vspace{.1cm}$


\textbf{Theorem $\mathbf{1}$}: In the original $\tilde{Q}$ model, the probability distributions remain invariant $\langle\psi|\prod_{i=1}^{n}E^{a_{i},\mathbf{x}_{i}}_{i}|\psi\rangle$ 
$=\langle\psi'|\prod_{j=1}^{n}E^{a_{j},\mathbf{x}'_{j}}_{j}|\psi'\rangle$ [Eq. (\ref{pro2})] if and only if Eqs. (\ref{rot1}) hold.

$\vspace{.1cm}$


\subsubsection{Proof for specific subset of unitary transformations}

Similar to the proof of Theorem $2$ in the main text, we take a specific subset of the unitary transformations (\ref{rot1}) such that $U_{1}(\theta_{1},\mathbf{k}_{1})\neq\openone, \hspace{.1cm} (\theta_{1}\neq0)$ but $U_{i\neq1}(\theta_{i},\mathbf{k}_{i})=\openone,\hspace{.1cm} (\theta_{i}=0)$, in which the sufficient and necessary conditions of Theorem $1$ can be summarized as follows:
\begin{eqnarray}\label{pro13}
&&\langle\psi|\prod_{i=1}^{n}E^{a_{i},\mathbf{x}_{i}}_{i}|\psi\rangle=
\langle\psi|U_{1}^{\dag}(\theta_{1},\mathbf{k}_{1})
\pi\left(U_{1}(\theta_{1},\mathbf{k}_{1})E^{a_{1},\mathbf{x}_{1}}_{1}U_{1}^{\dag}(\theta_{1},\mathbf{k}_{1})
\prod_{j=2}^{n}E^{a_{j},\mathbf{x}_{j}}_{j}\right)U_{1}(\theta_{1},\mathbf{k}_{1})|\psi\rangle,\hspace{.2cm}(\text{I}) \nonumber\\
&&\prod_{j=2}^{n}E^{a_{j},\mathbf{x}_{j}}_{j}U_{1}(\theta_{1},\mathbf{k}_{1})|\psi\rangle
=\pi\left(\prod_{j=2}^{n}E^{a_{j},\mathbf{x}_{j}}_{j}U_{1}(\theta_{1},\mathbf{k}_{1})\right)|\psi\rangle,\hspace{.2cm}(\text{II}) \hspace{1cm}\forall\,\theta_{1}, \hspace{.1cm}\mathbf{k}_{1}.
\end{eqnarray}
Similar Eq. as (\ref{pro13}-\textbf{II}) holds for $U_{1}^{\dag}(\theta_{1},\mathbf{k}_{1})$. In what follows, we use invariance of probability distributions under symmetry transformations and the postulates of AQCs model. The permutation relations (\ref{rot1}-$iii$) do not paly any role in the proof of Theorem $1$.

\emph{The proof of the sufficient condition} $\text{II}\Longrightarrow \text{I}$

Due to the definition of the permutation relations (\ref{per1}-\text{II}), the $U_{1} (U_{1}^{\dag})$ in the probability distribution (\ref{pro13}-\text{I}) can move to the left/right side of the projection operators $E^{a_{j},\mathbf{x}_{j}}_{j}, \hspace{.1cm} j=2,\cdots,n$ on the $|\psi\rangle$ and it is neutralized by $U_{1}^{\dag}(U_{1})$.
These indicates that the probability distribution remains invariant.

\emph{The proof of the necessary condition} $\text{I}\Longrightarrow \text{II}$

We know that postulate ($\mathbf{3}'$) holds for arbitrary subset of qudits. Here, we show similar permutation relations exist among unitary transformations and the projection operators. In particular, action of any arbitrary permutation of the first qudit unitary transformations and the rest of projection operators on the state vector $|\psi\rangle$ are equal.

We take an infinitesimal rotation/translation parameter $\delta\theta_{1}\ll 1$ for the first party, where the unitary transformation is given by $U_{1}(\delta\theta_{1},\mathbf{k}_{1})=\openone-i \delta\theta_{1}M_{1}(\mathbf{k}_{1})$ with $M_{1}(\mathbf{k}_{1})$ being the generator of the rotation/translation group around the direction $\mathbf{k}_{1}$. Therefore, the probability distribution (\ref{pro13}-\textbf{I}) is given by
\begin{eqnarray}\label{pro14}
&&\langle\psi|\prod_{i=1}^{n}E^{a_{i},\mathbf{x}_{i}}_{i}|\psi\rangle=
\langle\psi|\left[\openone +i \delta\theta_{1}M_{1}(\mathbf{k}_{1})\right]
\pi\big(\left[\openone -i \delta\theta_{1}M_{1}(\mathbf{k}_{1})\right]E^{a_{1},\mathbf{x}_{1}}_{1}\big.\\
&&\hspace{6cm}\big.\left[\openone +i \delta\theta_{1}M_{1}(\mathbf{k}_{1})\right]\prod_{j=2}^{n}E^{a_{j},\mathbf{x}_{j}}_{j}\big)\left[\openone -i \delta\theta_{1}M_{1}(\mathbf{k}_{1})\right]|\psi\rangle\nonumber\\
&&\hspace{1.9cm}=\langle\psi|\pi\big(\prod_{i=1}^{n}E^{a_{i},\mathbf{x}_{i}}_{i}\big)|\psi\rangle
+i\delta\theta_{1}\left[\langle\psi|M_{1}(\mathbf{k}_{1})\pi\big(\prod_{i=1}^{n}E^{a_{i},\mathbf{x}_{i}}_{i}\big)|\psi\rangle\right.\nonumber\\
&&\hspace{1.9cm}\left.-\langle\psi|
\pi\big(M_{1}(\mathbf{k}_{1})\prod_{j=1}^{n}E^{a_{j},\mathbf{x}_{j}}_{j}\big)|\psi\rangle
+\langle\psi|
\pi\big(E^{a_{1},\mathbf{x}_{1}}_{1}M_{1}(\mathbf{k}_{1})\prod_{j=2}^{n}E^{a_{j},\mathbf{x}_{j}}_{j}\big)|\psi\rangle
\right.\nonumber\\\nonumber
&&\hspace{3.3cm}-\left.\langle\psi|\pi\big(\prod_{j=1}^{n}E^{a_{j},\mathbf{x}_{j}}_{j}\big)M_{1}(\mathbf{k}_{1})|\psi\rangle\right]
+O((\delta\theta_{1})^{2})+\cdots,
\,\forall\hspace{.1cm}\mathbf{k}_{1}.
\end{eqnarray}
After discarding the second and higher order terms, we find
\begin{eqnarray}\label{pro23}
&&0=\langle\psi|M_{1}(\mathbf{k}_{1})\pi\big(\prod_{i=1}^{n}E^{a_{i},\mathbf{x}_{i}}_{i}\big)|\psi\rangle
-\langle\psi|\pi\big(M_{1}(\mathbf{k}_{1})\prod_{j=1}^{n}E^{a_{j},\mathbf{x}_{j}}_{j}\big)|\psi\rangle\nonumber\\
&& \hspace{.9cm}+\langle\psi|\pi\big(E^{a_{1},\mathbf{x}_{1}}_{1}M_{1}(\mathbf{k}_{1})\prod_{j=2}^{n}E^{a_{j},\mathbf{x}_{j}}_{j}\big)|\psi\rangle
-\langle\psi|\pi\big(\prod_{j=1}^{n}E^{a_{j},\mathbf{x}_{j}}_{j}\big)M_{1}(\mathbf{k}_{1})|\psi\rangle,
\hspace{.1cm}\forall\hspace{.1cm}\mathbf{k}_{1}.
\end{eqnarray}
The permutation operators in the second and third terms in Eq. (\ref{pro23}) can be used to show that the following equalities hold
\begin{eqnarray*}\label{}
&&\langle\psi|M_{1}(\mathbf{k}_{1})\pi\big(\prod_{i=1}^{n}E^{a_{i},\mathbf{x}_{i}}_{i}\big)|\psi\rangle
=\langle\psi|\pi\big(M_{1}(\mathbf{k}_{1})\prod_{j=1}^{n}E^{a_{j},\mathbf{x}_{j}}_{j}\big)|\psi\rangle\nonumber\\
&&\langle\psi|\pi\big(E^{a_{1},\mathbf{x}_{1}}_{1}M_{1}(\mathbf{k}_{1})\prod_{j=2}^{n}E^{a_{j},\mathbf{x}_{j}}_{j}\big)|\psi\rangle
=\langle\psi|\pi\big(\prod_{j=1}^{n}E^{a_{j},\mathbf{x}_{j}}_{j}\big)M_{1}(\mathbf{k}_{1})|\psi\rangle,
\hspace{.1cm}\forall\hspace{.1cm}\mathbf{k}_{1}.
\end{eqnarray*}
The above relation indicates that the symmetry generator $M_{1}(\theta_{1},\mathbf{k}_{1})$ can move to the right-(left)-hand side of the projection operators $E^{a_{j},\mathbf{x}_{j}}_{j}$ ($j\neq1$) when they act on the state $|\psi\rangle$. This relation is equivalent to
\begin{eqnarray}\label{pro24}
&&\langle\psi|U_{1}(\delta\theta_{1},\mathbf{k}_{1})\pi\big(\prod_{i=1}^{n}E^{a_{i},\mathbf{x}_{i}}_{i}\big)|\psi\rangle
=\langle\psi|\pi\big(U_{1}(\delta\theta_{1},\mathbf{k}_{1})\prod_{j=1}^{n}E^{a_{j},\mathbf{x}_{j}}_{j}\big)|\psi\rangle\nonumber\\
&&\langle\psi|\pi\big(E^{a_{1},\mathbf{x}_{1}}_{1}U_{1}(\delta\theta_{1},\mathbf{k}_{1})\prod_{j=2}^{n}E^{a_{j},\mathbf{x}_{j}}_{j}\big)|\psi\rangle
=\langle\psi|\pi\big(\prod_{j=1}^{n}E^{a_{j},\mathbf{x}_{j}}_{j}\big)U_{1}(\delta\theta_{1},\mathbf{k}_{1})|\psi\rangle,
\hspace{.1cm}\forall\hspace{.1cm}\mathbf{k}_{1},
\end{eqnarray}
for an arbitrary infinitesimal rotation/translation parameter $\delta\theta_{1}\ll 1$.

By applying another unitary transformations on the first party and repeating similar procedure as Eq. (\ref{pro23}), we arrive at a similar relation as Eq. (\ref{pro24}) for a finite unitary transformation $U_{1}(\theta_{1},\mathbf{k}_{1})$ as
\begin{eqnarray*}\label{}
&&\langle\psi|U_{1}(\theta_{1},\mathbf{k}_{1})\pi\big(\prod_{i=1}^{n}E^{a_{i},\mathbf{x}_{i}}_{i}\big)|\psi\rangle
=\langle\psi|\pi\big(U_{1}(\theta_{1},\mathbf{k}_{1})\prod_{j=1}^{n}E^{a_{j},\mathbf{x}_{j}}_{j}\big)|\psi\rangle\nonumber\\
&&\langle\psi|\pi\big(E^{a_{1},\mathbf{x}_{1}}_{1}U_{1}(\theta_{1},\mathbf{k}_{1})\prod_{j=2}^{n}E^{a_{j},\mathbf{x}_{j}}_{j}\big)|\psi\rangle
=\langle\psi|\pi\big(\prod_{j=1}^{n}E^{a_{j},\mathbf{x}_{j}}_{j}\big)U_{1}(\theta_{1},\mathbf{k}_{1})|\psi\rangle,
\hspace{.1cm}\forall\hspace{.1cm}\mathbf{k}_{1}.
\end{eqnarray*}
This equation says that the expectation value any permutation of $U_{1}(\theta_{1},\mathbf{k}_{1})$ and $\prod_{j=2}^{n}E^{a_{j},\mathbf{x}_{j}}_{j}$ are equal.
By summing on the projection operators of the first party and using the completeness relation $\sum_{a_{1}}E^{a_{1},\mathbf{x}_{1}}_{1}=\openone$, we find an expression independent of $E^{a_{1},\mathbf{x}_{1}}_{1}$ as:
\begin{eqnarray}\label{pro16}
\langle\psi|U_{1}(\theta_{1},\mathbf{k}_{1})\left(\prod_{j=2}^{n}E^{a_{j},\mathbf{x}_{j}}_{j}\right)|\psi\rangle
=\langle\psi|
\pi\left(U_{1}(\theta_{1},\mathbf{k}_{1})
\prod_{j=2}^{n}E^{a_{j},\mathbf{x}_{j}}_{j}\right)|\psi\rangle,\hspace{.3cm}\forall \hspace{.1cm}\mathbf{k}_{1},\theta_{1},
\end{eqnarray}
which indicates that the expectation values on the state $|\psi\rangle$ of any permutation of $U_{1}(\theta_{1},\mathbf{k}_{1})$ and $E^{a_{j},\mathbf{x}_{j}}_{j}$ are all equal. In the above relation, the $E^{a_{j},\mathbf{x}_{j}}_{j}$ ($j=2,\cdots,n$) project the state vector $\langle\psi|$ on engine-vectors of $E^{a_{j},\mathbf{x}_{j}}_{j}$ and the $U_{1}$ rotates $\langle\psi|$ in the Hilbert space $\mathcal{H}_{1}$.

We now replace the unitary transformation $U_{1}(\theta_{1},\mathbf{k}_{1})$ with sequence of three unitary transformations $U_{1}(\theta_{1},\mathbf{k}_{1})\rightarrow V_{1}(\upsilon_{1},\mathbf{v}_{1})U_{1}(\theta_{1},\mathbf{k}_{1})W_{1}(\omega_{1},\mathbf{w}_{1})$, the $V_{1}(\upsilon_{1},\mathbf{v}_{1})$ ($W_{1}(\omega_{1},\mathbf{w}_{1})$) is an arbitrary unitary transformation around the direction $\mathbf{v}_{i}$  ($\mathbf{w}_{1}$) with a finite values (angle/distance) $\upsilon_{1}$ ($\omega_{1}$) on the first qudit. The relation (\ref{pro16}) changes as:
\begin{eqnarray}\label{pro31}
&&\langle\psi|V_{1}(\upsilon_{1},\mathbf{v}_{1})
U_{1}(\theta_{1},\mathbf{k}_{1})W_{1}(\omega_{1},\mathbf{w}_{1})\prod_{j=2}^{n}E^{a_{j},\mathbf{x}_{j}}_{j}|\psi\rangle\\\nonumber
&&\hspace{4cm}=\langle\psi|
\pi_{V<U<W}\left(V_{1}(\upsilon_{1},\mathbf{v}_{1})
U_{1}(\theta_{1},\mathbf{k}_{1})W_{1}(\omega_{1},\mathbf{w}_{1})
\prod_{j=2}^{n}E^{a_{j},\mathbf{x}_{j}}_{j}\right)|\psi\rangle,
\end{eqnarray}
here, we used the index $V<U<W$ on the permutation operator $\pi_{V<U<W}$ to emphasis that $V_{1}(\upsilon_{1},\mathbf{v}_{1})$ ($W_{1}(\omega_{1},\mathbf{w}_{1})$) always appear in the left(right)-hand side of $U_{1}(\theta_{1},\mathbf{k}_{1})$. Due to the (\ref{pro31}), the $V_{1}$ ($W_{1}$) can move to left (right) hand side of sequence of projection operators and gives
\begin{eqnarray}\label{pro32}
\langle\psi_{V_{1}}|
U_{1}(\theta_{1},\mathbf{k}_{1})\prod_{j=2}^{n}E^{a_{j},\mathbf{x}_{j}}_{j}|\psi_{W_{1}}\rangle
=\langle\psi_{V_{1}}|\pi\big(U_{1}(\theta_{1},\mathbf{k}_{1})\prod_{j=2}^{n}E^{a_{j},\mathbf{x}_{j}}_{j}\big)|\psi_{W_{1}}\rangle,
\end{eqnarray}
where, $\langle\psi_{V_{1}}|=\langle\psi|V_{1}(\upsilon_{1},\mathbf{v}_{1})$ and $|\psi_{W_{1}}\rangle=W_{1}(\omega_{1},\mathbf{w}_{1})|\psi\rangle$. We can generate all elements of Hilbert space $\mathcal{H}_{1}$ by varying the rotation/trabslation parameter $\upsilon_{1}$ ($\omega_{1}$) and rotation/translation direction $\mathbf{v}_{1}$ ($\mathbf{w}_{1}$) of first qudit.

From Eq. (\ref{pro32}) we can conclude two points: ($1$) Expectation value of any arbitrary permutation of the $U_{1}$ and $E^{a_{j},\mathbf{x}_{j}}_{j},\, j=2,\cdots,n$ on the ket state $|\psi_{W_{1}}\rangle$ and the bra state $\langle\psi_{V_{1}}|$ are equal which indicates the state of the first party can be arbitrary. ($2$) The state vector of the first party does not play any role in determination of matrix elements rest of projection operators $E^{a_{j},\mathbf{x}_{j}}_{j}, \hspace{.1cm}j=2,\cdots,n$.



On other hand, we take summation on the first party measurement outputs $a_{1}=1,\cdots,d_{1}$ in postulate ($\mathbf{3}'$) and use completeness relation of projection operator $\sum_{a_{1}=1}^{d_{1}}E^{a_{1},\mathbf{x}_{1}}_{1}=\openone$ (postulate ($\mathbf{2}'$)), consequently, we find the permutation relations for subset of $E^{a_{j},\mathbf{x}_{j}}_{j}$ $j=2,\cdots,n$ as:
\begin{eqnarray}
\label{pro33}
E^{a_{2},\mathbf{x}_{2}}_{2}\cdots E^{a_{n},\mathbf{x}_{n}}_{n}|\psi\rangle
=\pi\left(E^{a_{2},\mathbf{x}_{2}}_{2}\cdots E^{a_{n},\mathbf{x}_{n}}_{n}\right)|\psi\rangle,
\hspace{.3cm}\forall\hspace{.2cm} \hspace{.1cm}a_{i},\hspace{.3cm}i=1,\cdots,n,
\end{eqnarray}
and, we rewrite the Eq. (\ref{pro32}) with $\omega_{1}=0$
\begin{eqnarray}
\label{pro34}
\langle\psi_{V_{1}}|
U_{1}(\theta_{1},\mathbf{k}_{1})\prod_{j=2}^{n}E^{a_{j},\mathbf{x}_{j}}_{j}|\psi\rangle
=\langle\psi_{V_{1}}|\pi\big(U_{1}(\theta_{1},\mathbf{k}_{1})\prod_{j=2}^{n}E^{a_{j},\mathbf{x}_{j}}_{j}\big)|\psi\rangle.
\end{eqnarray}

In the computational bases $\{|a_{1},\cdots,a_{n}\rangle|\hspace{.1cm}|a_{1},\cdots,a_{n}\rangle=\bigotimes_{i=1}^{n}|a_{i}\rangle_{\mathbf{x}_{i}}, \hspace{.1cm}\forall\hspace{.1cm}a_{i}=1,\cdots,d_{i}, \hspace{.1cm} i=1,\cdots,n\}$ and for the specific set of inputs
$\left\{\mathbf{x}_{i}, \hspace{.1cm}\forall\hspace{.1cm} i\right\}$, the projection operators take diagonal representation with elements of $0$ and $1$. This property says that Eq. (\ref{pro34}) for $n-1$ subsets of qubits ($2,\cdots,n$) is equivalent to permutation relation matrix (\ref{pro33}). Hence, the state of first qudit is an arbitrary state $_{\mathbf{x'}_{1}}\langle a_{1}|=$ $_{\mathbf{x}_{1}}\langle a_{1}|V_{1}(\upsilon_{1},\mathbf{v}_{1})$ which covers all elements of Hilbert space $\mathcal{H}_{1}$.

According to these outcomes, we can conclude that the following equality holds
\begin{eqnarray*}\label{pro35}
U_{1}(\theta_{1},\mathbf{k}_{1})\prod_{j=2}^{n}E^{a_{j},\mathbf{x}_{j}}_{j}|\psi\rangle
=\pi\left(U_{1}(\theta_{1},\mathbf{k}_{1})
\prod_{j=2}^{n}E^{a_{j},\mathbf{x}_{j}}_{j}\right)|\psi\rangle,
\end{eqnarray*}
this relation is equal to the relation (\ref{pro13}-II). Then, the necessary condition is proved.

\emph{Remark 1}: Merging Eqs. (\ref{pro33}) and (\ref{pro34}) can be realized by applying postulate $(\textbf{2}')$, which indicates $E^{a_{i},\mathbf{x}_{i}}_{i}E^{a_{i},\mathbf{x}_{i}}_{i}=E^{a_{i},\mathbf{x}_{i}}_{i}$, to rewriting Eq. (\ref{pro34}) as:
\begin{eqnarray*}\label{}
\langle\psi_{V_{1}}|\prod_{j=2}^{n}E^{a_{j},\mathbf{x}_{j}}_{j}
U_{1}(\theta_{1},\mathbf{k}_{1})\prod_{j=2}^{n}E^{a_{j},\mathbf{x}_{j}}_{j}|\psi\rangle
=\langle\psi_{V_{1}}|\prod_{j=2}^{n}E^{a_{j},\mathbf{x}_{j}}_{j}
\pi\big(U_{1}(\theta_{1},\mathbf{k}_{1})\prod_{j=2}^{n}E^{a_{j},\mathbf{x}_{j}}_{j}\big)|\psi\rangle,
\end{eqnarray*}
the action of the projection operators $E^{a_{j},\mathbf{x}_{j}}_{j}, \, j=2,\cdots,n$ on the bra state reduces the above equation to
\begin{eqnarray*}\label{}
&&\left(_{\mathbf{x'}_{1}}\langle a_{1}|\right)\left(_{\mathbf{x}_{i}}\langle a_{i}|\bigotimes_{i=2}^{n}\right)
U_{1}(\theta_{1},\mathbf{k}_{1})\prod_{j=2}^{n}E^{a_{j},\mathbf{x}_{j}}_{j}|\psi\rangle\\
&&\hspace{5cm}=\left(_{\mathbf{x'}_{1}}\langle a_{1}|\right)\left(_{\mathbf{x}_{i}}\langle a_{i}|\bigotimes_{i=2}^{n}\right)
\pi\big(U_{1}(\theta_{1},\mathbf{k}_{1})\prod_{j=2}^{n}E^{a_{j},\mathbf{x}_{j}}_{j}\big)|\psi\rangle,
\end{eqnarray*}
this equation is correct for each element of the Hilbert space $\mathcal{H}=\bigotimes_{i=1}^{n}\mathcal{H}_{i}$ in the computational bases with values of zero or $_{\mathbf{x'}_{1}}\langle a_{1}|U_{1}(\theta_{1},\mathbf{k}_{1})|a_{1}\rangle_{\mathbf{x}_{1}}$. Therefore, it can be represented in the matrix form (\ref{pro13}-II).

\subsubsection{Derivation of the permutation relations (\ref{rot1}-$ii$) and (\ref{rot1}-$iv$)}\label{der}

As indicated in the previous section, we have proved a special case of the theorem $1$ for a specific set of the unitary transformations, which will help us now to prove it in the general case. Eq. (\ref{pro13}-II) is useful to obtain the permutation relations (\ref{rot1}-$ii$) and (\ref{rot1}-$iv$).

As pointed out in Eq. (\ref{pro7}) of the main text, we can expand the unitary transformation $U_{1}(\theta_{1},\mathbf{k}_{1})$ in Eq. (\ref{pro13}-II) and use the completeness relation to find
\begin{eqnarray*}\label{pro80}
\prod_{j=2}^{n}E^{a_{j},\mathbf{x}_{j}}_{j}E^{a_{1},\mathbf{k}_{1}}_{1}|\psi\rangle
=\pi\left(\prod_{j=2}^{n}E^{a_{j},\mathbf{x}_{j}}_{j}E^{a_{1},\mathbf{k}_{1}}_{1}\right)|\psi\rangle,
\hspace{.2cm} \forall \hspace{.2cm} a_{1}=1,\cdots,d_{1}.
\end{eqnarray*}
Similar to the main text, we replace the permutation relation of postulate ($\mathbf{3}'$) with Eq. (\ref{pro8}) and repeat a similar scenario for other projection operators and extend this postulate to arbitrary sets of measurement inputs,
\begin{eqnarray}\label{pro90}
\prod_{j=1}^{n}E^{a_{j},\mathbf{k}_{j}}_{j}|\psi\rangle
=\pi\left(\prod_{j=1}^{n}E^{a_{j},\mathbf{k}_{j}}_{j}\right)|\psi\rangle, \hspace{.2cm} \forall \hspace{.1cm} a_{j},\hspace{.3cm} i=1,\cdots,n.
\end{eqnarray}
Thus, this way we show that postulate ($\mathbf{3}'$) holds for any set of measurement inputs.

This relation helps extend postulate ($\mathbf{3}'$) to every permutation relations of arbitrary subset of the projection operators and the unitary transformations. To show this, we apply for one of the parties ($i$th subsystem for example) Eq. (\ref{pro90}) for every one of measurement outputs $a_{i}=1,\cdots,d_{i}$, which gives the sequence
\begin{eqnarray}\label{pro85}
\left[\prod_{j=1, \hspace{.1cm j\neq i}}^{n}E^{a_{j},\mathbf{k}_{j}}_{j}\right]E^{a_{i}=1,\mathbf{k}_{i}}_{i}|\psi\rangle
&=&\pi\left(\left[\prod_{j=1, \hspace{.1cm j\neq i}}^{n}E^{a_{j},\mathbf{k}_{j}}_{j}\right]E^{a_{i}=1,\mathbf{k}_{i}}_{i}\right)|\psi\rangle,\nonumber\\
\left[\prod_{j=1, \hspace{.1cm j\neq i}}^{n}E^{a_{j},\mathbf{k}_{j}}_{j}\right]E^{a_{i}=2,\mathbf{k}_{i}}_{i}|\psi\rangle
&=&\pi\left(\left[\prod_{j=1, \hspace{.1cm j\neq i}}^{n}E^{a_{j},\mathbf{k}_{j}}_{j}\right]E^{a_{i}=2,\mathbf{k}_{i}}_{i}\right)|\psi\rangle,\\\nonumber
&\cdots &\\\nonumber
\left[\prod_{j=1, \hspace{.1cm j\neq i}}^{n}E^{a_{j},\mathbf{k}_{j}}_{j}\right]E^{a_{i}=d_{i},\mathbf{k}_{i}}_{i}|\psi\rangle
&=&\pi\left(\left[\prod_{j=1, \hspace{.1cm j\neq i}}^{n}E^{a_{j},\mathbf{k}_{j}}_{j}\right]E^{a_{i}=d_{i},\mathbf{k}_{i}}_{i}\right)|\psi\rangle.
\end{eqnarray}
As indicated in the main text, the unitary transformations can be expanded as liner combination of the projection operators $U_{i}(\theta_{i},\mathbf{k}_{i})=\sum_{a_{i}=1}^{d_{i}}e^{ic(\theta_{i},\mathbf{k}_{i},a_{i})} E^{a_{i},\mathbf{k}_{i}}_{i}$, for some real $c(\theta_{i},\mathbf{k}_{i},a_{i})$. Thus, form Eq. (\ref{pro85}) we obtain:
\begin{eqnarray*}\label{pro86}
\prod_{j=1, \hspace{.1cm j\neq i}}^{n}E^{a_{j},\mathbf{k}_{j}}_{j}U_{i}(\theta_{i},\mathbf{k}_{i})
=\pi\left(\prod_{j=1, \hspace{.1cm j\neq i}}^{n}E^{a_{j},\mathbf{k}_{j}}_{j}U_{i}(\theta_{i},\mathbf{k}_{i})\right)|\psi\rangle.
\end{eqnarray*}
We continue this procedure to generate the permutation relations between a subset  $I$ of the unitary transformations and its disjoint subset $J$ of the projection operators as
\begin{eqnarray*}\label{pro87}
\prod_{i\in I}U_{i}(\theta_{i},\mathbf{k}_{i}\prod_{j\in J}E^{a_{j},\mathbf{x}_{j}}_{j}|\psi\rangle
=\pi\left(\prod_{i\in I}U_{i}(\theta_{i},\mathbf{k}_{i})\prod_{j\in J}E^{a_{j},\mathbf{x}_{j}}_{j}\right)|\psi\rangle,
\hspace{.1cm}\forall \hspace{.1cm}\mathbf{k}_{i}, \,\theta_{i}, \,a_{j},\, i,\,j\in\{1,...,n\},
\end{eqnarray*}
where $I\subset\{1,...,n\},\, J\subset\{1,...,n\},\hspace{.1cm}\, I\cap J=\varnothing$. It is clear that $I$ ($J$) can range from the empty set $\varnothing$ to the full set $\{1,...,n\}$, which encompass the permutation relations of the projection operators in
postulate ($\mathbf{3}'$) up to the permutation relations of the unitary transformations of Eq. (\ref{rot1}-$ii$).

$\vspace{.1cm}$

\emph{Remark 2}: Here we have derived the permutation relations of the unitary transformations $U_{i}$ as a consequence of the invariance of the permutation relations of the projection operators under symmetry transformations. Note that violation of Eq. (\ref{rot1}-$ii$) may imply instantaneous communications. Specifically, two spacelike parties $i$ and $j$ synchronize their time such that if the $i$ does not perform any action, it means communication ``$0$" to party $j$, and otherwise it means ``$1$". In particular, this party performs nothing $U_{i}=\openone$ or it performs a sequence of the unitary transformations $U^{\dagger}_{i}$ and $U_{i}$ in the times $t-\delta t$ and $t+\delta t$ ($\delta t\ll 1$), respectively. Party $j$ performs the unitary transformation $U_{j}$ in the time $t$. Now from $U_{i}U_{j}U^{\dagger}_{i}|\psi\rangle=\tilde{U}_{j}|\psi\rangle\neq U_{j}|\psi\rangle, \hspace{.2cm}(i\neq j)$, we see that if party $i$ performs nothing or a sequence of the unitary transformations, then the state of party $j$ is, respectively, given by $U_{j}|\psi\rangle$ or $\tilde{U}_{j}|\psi\rangle$. Due to commuting relations of unitary transformations \cite{Gre}, it is possible that the reduce density matrices of party $j$, $\rho_{j}$ and $\tilde{\rho}_{j}$, become different $\rho_{j}\neq\tilde{\rho}_{j}$ for some types of unitary transformations $U_{i}$ and $U_{j}$. In particular, party $j$ compares statistic of outcomes $a_{j}=1,\cdots,d_{j}$ in density matrices $\rho_{j}$ and $\tilde{\rho}_{j}$, and concludes messages $0$ and $1$, respectively.



\subsubsection{Proof of Theorem $1$ in general cases}

A) Sufficient condition: If Eqs. (\ref{rot1}) hold, then the probability distributions remain invariant [Eq. (\ref{pro2})].

Due to Eq. (\ref{rot1}-\emph{iv}), each of the unitary transformations $U_{j}$ ($U_{j}^{\dag}$) in the middle term of the probability distribution (\ref{pro2}), $\prod_{j=1}^{n}U_{j}(\theta_{j},\mathbf{k}_{j})E^{a_{j},\mathbf{x}_{j}}_{j}U_{j}^{\dag}(\theta_{j},\mathbf{k}_{j})$, can move to the left-(right-)hand side of the projection operators $E^{a_{t},\mathbf{x}_{t}}_{t}, \, t\neq j$, which can be rearranged as (\ref{pro2}).
\begin{eqnarray*}\label{}
\langle\psi|\prod_{i=1}^{n}U_{i}^{\dagger}(\theta_{i},\mathbf{k}_{i})
\left(\prod_{r=1}^{n}U_{r}(\theta_{r},\mathbf{k}_{r})\right)
\left(\prod_{j=1}^{n}E^{a_{j},\mathbf{x}_{j}}_{j}\right)
\left(\prod_{s=1}^{n}U_{s}^{\dag}(\theta_{s},\mathbf{k}_{s})\right)
\prod_{k=1}^{n}U_{k}(\theta_{k},\mathbf{k}_{k})|\psi\rangle.
\end{eqnarray*}
As pointed out in the main text, $U_{j}$ ($U_{j}^{\dag}$) should always appear on the left-(right-)hand side of $E_{j}$.

We now use the permutation relations (\ref{rot1}-\emph{ii}) and $\langle\psi'|=\langle\psi|\prod_{i=1}^{n}U_{i}(\theta_{i},\mathbf{k}_{i})
=\langle\psi|\pi\left(\prod_{i=1}^{n}U_{i}(\theta_{i},\mathbf{k}_{i})\right)$ to neutralize each $U_{i}^{\dag}$ with $U_{i}$ on the left-(right-)hand side, which indicates that the probability distributions remain invariant.

B) Necessary condition: If the probability distributions remain invariant [Eq. (\ref{pro2})], then Eqs. (\ref{rot1}) hold.

In Sec. \ref{D}-2 we have derived Eqs. (\ref{rot1}-$ii$) and (\ref{rot1}-$iv$) as results of Eq. (\ref{pro2}). Hence, this completes the proof of Theorem $1$. \hfill $\blacksquare$

\subsection{Proof that the arbitrary permutation relations of the unitary transformations are sufficient to reduce the set $\tilde{Q}$ to standard quantum mechanics} \label{E}

We take the sets of unitary transformations (\ref{rot1}-$iii$) and replace $U_{i}(\theta_{i},\mathbf{k}_{i})\rightarrow U_{i}(\theta_{i},\mathbf{k}_{i}) V_{i}(\nu_{i},\mathbf{v}_{i}), \hspace{.2cm}\forall\hspace{.1cm} \mathbf{v}_{1}, \nu_{1}$:
\begin{eqnarray}\label{rot11}
&&\prod_{i=1}^{n}U_{i}(\theta_{i},\mathbf{k}_{i})V_{i}(\nu_{i},\mathbf{v}_{i})|\psi\rangle
=\pi\left(\prod_{i=1}^{n}U_{i}(\theta_{i},\mathbf{k}_{i})V_{i}(\nu_{i},\mathbf{v}_{i})\right)|\psi\rangle\\\nonumber
&&\hspace{3cm}=\pi_{U<V}\left(\prod_{i=1}^{n}U_{i}(\theta_{i},\mathbf{k}_{i})\prod_{j=1}^{n}V_{j}(\nu_{j},\mathbf{v}_{j})\right)|\psi\rangle,
\hspace{.5cm}\forall\hspace{.1cm} \theta_{i}, \mathbf{k}_{i}, \nu_{i}, \mathbf{v}_{i}, \hspace{.2cm} i=2,\cdots,n.
\end{eqnarray}
Similar to the previous section, here the index ``$V<U$" on the permutation operator $\pi_{V<U}$ indicates that $U_{i}(\theta_{i},\mathbf{k}_{i})$ always appears on the left-hand side of $V_{i}(\nu_{i},\mathbf{v}_{i})$.

We now re-describe the permutation relations (\ref{rot11}) as an arbitrary permutation of $U_{i}(\theta_{i},\mathbf{k}_{i}),\hspace{.1cm}i=2,\cdots,n$ on the state vector $|\Psi\rangle=\prod_{j=1}^{n}V_{j}(\nu_{j},\mathbf{v}_{j})|\psi\rangle$:
\begin{eqnarray*}\label{rot14}
&&\prod_{i=1}^{n}U_{i}(\theta_{i},\mathbf{k}_{i})|\Psi\rangle
=\pi\left(\prod_{i=1}^{n}U_{i}(\theta_{i},\mathbf{k}_{i})\right)|\Psi\rangle,
\hspace{.5cm}\forall\hspace{.1cm} \theta_{i}, \mathbf{k}_{i}, \nu_{i}, \mathbf{v}_{i}, \hspace{.2cm} i=2,\cdots,n.
\end{eqnarray*}
The state $|\Psi\rangle$ plays the same role as $|\psi\rangle$ in Eq. (\ref{rot11}). Each element of the Hilbert space $\mathcal{H}=\bigotimes_{i=1}^{n}\mathcal{H}_{i}$ can be generated by variation of the rotation/translation parameter $\nu_{i}$ around direction $\mathbf{v}_{i}$, $\left\{|\Psi\rangle||\Psi\rangle=\prod_{i=1}^{n}V_{i}(\nu_{i},\mathbf{v}_{i})|\psi\rangle,\hspace{.1cm}\forall\hspace{.1cm} \nu_{i}, \mathbf{v}_{i}, \hspace{.1cm} i=1,\cdots,n \right\}$. Hence, a general form of Eq. (\ref{rot11}) is as follows
\begin{eqnarray}\label{rot16}
\prod_{i=1}^{n}U_{i}(\theta_{i},\mathbf{k}_{i})|\Psi\rangle
=\pi \left[\prod_{i=1}^{n}U_{i}(\theta_{i},\mathbf{k}_{i})\right]|\Psi\rangle, \hspace{.3cm}\forall \hspace{.1cm}|\Psi\rangle\in\left\{|\Psi\rangle\right\},
\hspace{.1cm} \theta_{i}, \mathbf{k}_{i},\hspace{.2cm} i=1,\cdots,n.\nonumber
\end{eqnarray}
This indicates that the commutator of every two unitary transformations on any vector of the Hilbert space $\mathcal{H}$ is vanishes,
\begin{eqnarray*}\label{}
\left[U_{i}(\theta_{i},\mathbf{k}_{i}),U_{j}(\theta_{j},\mathbf{k}_{j})\right]|\Psi\rangle=0,\hspace{.1cm}\forall\hspace{.1cm}\theta_{i}, \mathbf{k}_{i},\hspace{.1cm} \theta_{j}, \mathbf{k}_{j}, \hspace{.2cm} i\neq j=1,\cdots,n,
\end{eqnarray*}
which is equivalent to vanishing the commutation relations of the unitary transformations,
$$\left[U_{i}(\theta_{i},\mathbf{k}_{i}),U_{j}(\theta_{j},\mathbf{k}_{j})\right]=0.$$

To find the commutation relation of the projection operators, we again expand the unitary transformations as a liner combination of the projection operators, $U_{i}(\theta_{i},\mathbf{k}_{i})=\sum_{a_{i}}c(\theta_{i},\mathbf{k}_{i},a_{i}) E^{a_{i},\mathbf{k}_{i}}_{i}$, where $c(\theta_{i},\mathbf{k}_{i},a_{i})$ are the expansion coefficients which are the eigenvalues of unitary matrix with modulus $1$. Hence, the commutation relation reduces to
\begin{eqnarray}
\label{pro-}
\sum_{a_{i},a_{j}}c(\theta_{i},\mathbf{k}_{i},a_{i})c(\theta_{j},\mathbf{k}_{j},a_{j})
\left[E^{a_{i},\mathbf{k}_{i}}_{i},E^{a_{j},\mathbf{k}_{j}}_{j}\right]=0,\hspace{.2cm}\forall\hspace{.1cm} \theta_{i},\mathbf{k}_{i},\hspace{.1cm} \theta_{j},\mathbf{k}_{j}, \hspace{.2cm} i\neq j=1,\cdots,n.
\end{eqnarray}
Now noting to postulate ($\mathbf{2}'$), we remind that the projection operators $E^{a_{i},\mathbf{k}_{i}}_{i}, \hspace{.2cm} i=1,\cdots,n$, represent a complete set of independent and orthonormal basis operators, Eq. (\ref{pro}) implies $\left[E^{a_{i},\mathbf{k}_{i}}_{i},E^{a_{j},\mathbf{k}_{j}}_{j}\right]=0, \hspace{.2cm} i\neq j$. This indicates that postulate ($\mathbf{3}'$) for the $\tilde{Q}$ model reduces to postulate ($\mathbf{3}$) for standard quantum mechanics.

\end{document}